# Unravelling 2,4-D – biochar interactions by molecular dynamics: adsorption modes and surface functionalities

*Rosie Wood,[1] Ondřej Mašek,[2] and Valentina Erastova[1]\**

[1] School of Chemistry, University of Edinburgh, Joseph Black Building, David Brewster Road, King's Buildings, Edinburgh EH9 3FJ, UK

[2] UK Biochar Research Centre, School of GeoSciences, University of Edinburgh, Crew Building, Alexander Crum Brown Road, King's Buildings, Edinburgh EH9 3FF, UK

\* valentina.erastova@ed.ac.uk

**ABSTRACT**: We report a molecular dynamics investigation of 2,4-dichlorophenoxyacetic acid (2,4-D) adsorption at the aqueous-biochar interface using experimentally constrained woody biochar models representative of softwood-derived biochars produced at 400, 600 and 800 °C. The models reproduce experimental descriptors (H/C, O/C, aromaticity, true density, and surface functionality) of their experimental counterparts, and simulations enable calculation of adsorption isotherms that align with available experimental measurements. Our results reveal that 2,4-D$^-$ uptake is governed by a synergy of three interaction classes: (i) $\pi - \pi$ and $\pi -$ Cl contacts with graphitic domains with either parallel or perpendicular alignments, (ii) polar interactions including H-bonding to surface –OH and other oxygen-containing groups, and (iii) Na$^+$-mediated cation bridging that links 2,4-D$^-$ anion to surface oxygens, that would have an increasing relevance for biochars near or above the pH at point of zero charge. Notably, we found that low-temperature produced biochars, which retain higher densities of surface O functionalities, exhibit higher adsorption per unit surface area due to cooperative polar interactions alongside $\pi - \pi$ binding, whereas medium-to-high temperature biochars rely more on $\pi - \pi$ and cation-bridging mechanisms. The distinct adsorption distances measured emphasize surface heterogeneity and porosity. Taken together, these atomistic insights corroborate experimental observations and yield actionable guidance for the rational design of biochars for remediation of anionic herbicides, highlighting how surface functionality and solution chemistry can be tuned to optimize sorption. Our approach provides a general framework to interrogate pollutant-biochar interactions and to inform remediation strategies.

# 1. Introduction

Herbicides are indispensable in modern agriculture and urban landscaping for controlling unwanted vegetation. However, their chemical and photochemical stability hinders both biodegradation and removal by conventional water-treatment processes (e.g., advanced oxidation, adsorption).[1,2] This leads to their accumulation in soils, leaching into water streams, which then adversely affects the environment, exposing non-targeted species, including microbial communities, plants, animals and humans.[3,4] Therefore, in addition to efforts to minimise herbicides release into ecosystem, it is essential to develop an efficient, inexpensive and readily accessible technology to remove them from the environment.

2,4-Dichlorophenoxyacetic acid (2,4-D) is the first commercial herbicide, developed in 1940s, and still among the most widely used systemic herbicides worldwide, while its persistence and mobility in the environment have been linked to detrimental effects on aquatic ecosystems and potential endocrine disruption in mammals.[5] With a pKa of 2.7, it exists in the environment as an anion and is highly soluble in water (900 mg L$^{-1}$). This means that it is poorly retained by soils and is frequently found in the groundwater streams.[6] A range of remediation techniques has been explored for 2,4-D removal, including precipitation and sedimentation, flotation, ion-exchange resins, electrochemical oxidation and biodegradation, yet sorption is among the most accessible and effective methods.[2,7–14] The key to success is in the adsorbent material used, and many have been explored, including polymers,[7] MOFs,[8] layered minerals such as bentonite clay[12] and layered double hydroxides,[11] and activated carbon.[9,13,14] While commercially available activated carbons are frequently investigated in laboratory settings and provide most promising results, those materials are unaffordable both cost-wise and logistically to be used in large, contaminated areas, which often lay in low-income countries and remote communities. Therefore, biochar – an inexpensive, carbon-rich product of biomass pyrolysis – has emerged as a promising sorbent due to its high porosity, surface area, and tuneable surface chemistry.[10,13,15] Biochar production conditions (feedstock type, pyrolysis temperature, residence time) together with pre- or post-treatments (activation, oxidation, mineral loading) govern its textural properties and the density of surface functional groups. By controlling these parameters, it is possible to create engineered biochars tailored to specific contaminants and environmental conditions. Yet, decisions on biochar preparation for desired applications often rely on trial-and-error rather than knowledge of underlying chemical structure and functionality.

To this end, over the last decade, numerous experimental studies of biochar for the adsorption of 2,4-D have emerged, each demonstrating remarkable performance of the specific reported biochar [10,13,15] While some studies report no correlation between 2,4-D uptake and biochar chemical descriptors (mol% of C, H and O), that the key factor is the biochar's specific surface area;[13] others suggest that presence of specific functional groups (such as oxygen containing hydroxyl, carbonyl, carboxyl and also phosphoric acid and ammonia modifications) elevates adsorption capacities per unit surface area when compared to highly graphitised material.[10,15] These most recent works by Liu *et al.* and Tang *et al.* are implicating a critical role for polar and hydrogen-bonding interactions in addition to π – π stacking of graphitic sheets. However, the structural and chemical heterogeneity of real biochars makes it difficult to dissect these mechanisms experimentally

In this work, we apply atomistic molecular dynamics simulations to study adsorption mechanisms of 2,4-D at the aqueous biochar interface, using realistic experimentally-constrained biochar models. Our biochar molecular models, represent softwood-derived biochars produced at 400, 600 and 800 °C, and were constructed to match target experimental H/C and O/C molar ratios, aromaticity, functional groups, true density and morphology.[16] Based on molecular dynamics simulations, we report adsorption isotherms that quantitatively agree with reported experimental data and, through detailed geometric and coordination analyses, identify the interplay of (i) $\pi - \pi$ and $\pi - Cl$ interactions, (ii) hydrogen bonding to surface O-groups, and (iii) cation-bridging complexes with $Na^+$. These findings not only provide insights into experimental observations, but also deliver practical guidelines for the rational design of engineered biochars for remediation of anionic herbicide pollutants and supports sustainable agricultural practices and environmental protection

## 2. Methodology

We study adsorption of 2,4-D onto three biochars, representative of those produced from softwood feedstock at temperatures of 400 °C, 600 °C and 800 °C Thereafter, the biochar models are regarded as BC400, BC600 and BC800, respectively. Biochar surfaces were kept neutral, representative of the system at pH of point zero charge, which is at pH≅7-7.5 for woody biochars.[17] Since pKa of 2,4-D is 2.64,[18] in natural environments, it is present in anionic form 2,4-D$^-$. For each biochar system, we introduce 2,4-D$^-$ in water with the initial concentrations of 0.1, 0.5, 1, 1.25, 1.75, 2, 2.5 and 5 mol L$^{-1}$. These systems are charge-balanced with Na$^+$ cations, overall making them neutral. Although the concentrations used in this work would appear significantly higher than those found in environmental waters (~10× higher), this ratio of biochar's surface area to the number of pollutant molecules allows for simulations of 10s of nanoseconds and of sizes of 10s of nanometers to obtain statistically representative sampling. This, therefore, provides more reliable statistics for further analysis. Upon set-up, each system was run in the isothermal-isobaric ensemble, characteristic of realistic experimental conditions, i.e., atmospheric pressure and room temperature. The simulations were first equilibrated (30+ ns), then simulated for another 20 ns from which the statistics were collected and analysis performed. Our protocol is shown schematically in **Figure 1**, while technical details of the set-up and analysis are given below.

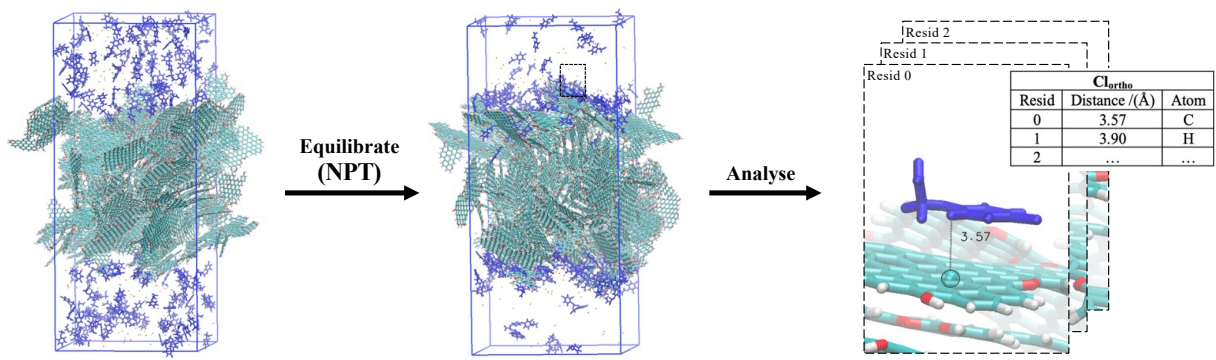

*Figure 1.* *A schematic guide to our equilibration and analysis procedure. Biochar atoms are shown in cyan (C), red (O) and white (H), 2,4-D$^-$ molecules are shown in blue and Na$^+$ ions are shown in orange, water molecules are not shown for clarity.*

### 2.1 Construction of Biochar and 2,4-D$^-$ Systems

In this work, we benefit from our previously developed and well-characterised biochar models.[16] These models have been carefully matched to the target experimental characteristics of woody biochars produced at a range of temperatures, for full details on the collection of experimental target parameters, the development of realistic biochar models and to download the biochar structures for simulations, we recommend referring to works by Wood *et al.*.[16,19]

The biochar models were first developed as bulk, i.e., periodic in *x*, *y* and *z*-directions; then two surfaces in *xy*-plane (top and bottom of the biochar) were created by expanding the simulation box along the *z*-axis while keeping the biochar material in the centre of the simulation box. These surfaces were solvated and equilibrated to allow any surface groups to arrange in the new environment. After equilibration, the biochar system was characterised, ensuring match to the experimental descriptors. To the equilibrated, surface-exposed solvated biochar systems, we added 2,4-D$^-$, charge-balanced by Na$^+$, to achieve desired starting concentrations. See **Table S1.2** (SI) for the number of Na$^+$, 2,4-D$^-$ and water molecules in each of the 24 systems simulated.

## 2.2 Choice of molecular dynamics engine and force field

All simulations were performed using GROMACS 2022 engine.[20] Unless otherwise specified, the set-up and analysis are performed with the inbuilt GROMACS tools. All of the molecular structures were assigned with OPLS-AA force field.[21,22] This force field was chosen for biochar systems due to its versatility and effectiveness in reproducing the properties of organic systems. Using OPLS-AA also allowed us to benefit from LigParGen[23] and PolyParGen[24] tools to assign the force fields to molecular structures. Single point charge (SPC) water and Na$^+$ cation parameters within OPLS-AA were used.

## 2.3 Simulation Details

### 2.3.1 Energy minimisation

The first step in the simulation of a newly prepared system is energy minimisation, to resolve any high forces. In all cases, it was carried out using the steepest descent algorithm, and proceeded until the maximum force on each atom was less than 500 kJ mol$^{-1}$ nm$^{-1}$. Each energy minimisation simulation used periodic boundary conditions in *x*, *y* and *z*, a 1.4 nm van der Waals (vdW) cut-off and Particle-Mesh-Ewald (PME) electrostatics.

### 2.3.2 Molecular dynamics simulation in isothermal-isobaric ensemble

To simulate the biochar surface and adsorption of 2,4-D$^-$, we performed a long molecular dynamics simulation (50+ ns) in the isothermal-isobaric ensemble (NPT), with a timestep of 2 fs, constraining H-bonds. For these simulations, we used periodic boundary conditions in *x*, *y* and *z*, PME electrostatics, a 1.4 nm vdW cut-off and a Velocity-rescale thermostat at 300 K with a 0.1 ps coupling constant. We applied semi-isotropic (*xy*, *z*) pressure coupling using a Berendsen barostat with a 1 ps coupling constant and reference pressure of 1 bar. The length of simulation was determined based on the time needed for the system to converge to the equilibrium state (30+ ns), after which the simulation was run for an additional 20 ns to obtain statistically significant samples for the analysis.

### 2.3.3 Assessment of the convergence of simulation

During the simulation, we monitored each system with Root Mean Square Deviation (RMSD) for the most mobile subpart of the system, i.e., 2,4-D⁻ molecules, monitoring the positions over time against the start of the simulation. The plateau in the RMSD indicates that the systems have equilibrated (**Figure S2.1**). We then also validated the equilibration using *DynDen*,[25] which allows us to check pairwise cross-correlation of all components independently and all frames in the simulation (**Figure S2.2**, **S2.3**). After equilibration was achieved, the simulations were extended for an additional 20 ns. This part of trajectory was then used as average over this time, reporting standard deviations.

### 2.3.4 Data sharing

The files used in this work: output structure files, topology files, force field files and MD input parameter files in GROMACS format, are freely available on our group's GitHub: github.com/Erastova-group/24D_biochar, DOI: 10.5281/zenodo.17990328.

## 2.4 Analysis

### 2.4.1 Characterisation of surface-exposed biochar material

Biochar is a porous material, and therefore, to calculate its true density, we must first exclude the unoccupied volume within the material. This free volume was calculated using the GROMACS *freevolume* tool, using a probe radius of 0.13 nm (representative of the He atom). The true density of biochar, $\rho_{solid}$, is then calculated as:

$$\rho_{solid} = \frac{\rho_{system}}{1 - V_{0.13}},$$

where $\rho_{system}$ is the density of the whole biochar system, and $V_{0.13}$ is the free volume of the system using a probe radius of 0.13 nm.

The exposed surface areas of biochar were characterised through their solvent accessible surface areas (SASAs), obtained with GROMACS *sasa* tool, using a probe radius of 0.18 nm (corresponding to N₂ gas). Due to variations in the *xy*-dimensions of each of our systems, we report normalised SASAs, $nSASA$, calculated as follows:

$$nSASA = \frac{SASA_{0.18}}{2 \times A_{xy}},$$

where $SASA_{0.18}$ is SASA of the system measured with a probe radius of 0.18 nm, and $A_{xy}$ is the $xy$-cross-sectional area of the simulation box.

To identify surface-exposed biochar atoms upon each biochar molecular model, we used the Pytim library.[26] This library is built on top of the MDAnalysis library[27,28] and is used for interfacial analysis in molecular dynamics simulations. We used the *ITIM* module and probe radius ($\alpha$ value) of 0.13 nm (corresponding to He gas, for consistency with density calculations) to identify the surface and assign atoms corresponding to the outer layers of biochar molecular models. For each system, we save IDs surface-exposed atoms for use in further analysis of adsorption, including counts of the total number of C, H and O atoms to calculate the atomic compositions of biochar surfaces.

### 2.4.2 Isotherms

One of the most straightforward and commonly used analyses to quantify adsorption is linear density perpendicular to the surface of the material. To this end, we used MDAnalysis' *lineardensity* module to calculate linear density profiles as an average across every 100 timesteps in the last 20 ns of our converged trajectories.

Since biochars feature non-planar surfaces, linear densities are not ideal for the quantification of adsorption. To obtain isotherms, for each biochar and concentration, we counted the number of 2,4-D⁻ molecules with their centre of mass (COM) within 0.6 nm from the biochar surface. The distance of 0.6 nm was chosen as the highest probability peak (**Figure S3.3**) between the COM of 2,4-D⁻ and aromatic rings of surface-forming biochar moieties. We then also computed the number of 2,4-D⁻ molecules left in solution, and therefore its equilibrium concentration. This is calculated for each 100 timesteps of the equilibrated trajectory, reporting the average value and standard deviation.

### 2.4.3 Adsorption Mechanism

To gain insights into the mechanisms through which 2,4-D⁻ adsorbs to biochar, for each 2,4-D⁻ molecule in our system, we report distances between the functional groups and the closest biochar atoms. The groups are as follows: COM of aromatic ring, chlorine atoms, and COM of carboxylic group. We recorded the ID of the nearest biochar surface atom, its type (C, H or O) and the distance between it and the functional group. After repeating this for each timestep, we report how the distances between each 2,4-D⁻ functional groups and the biochar surface varied. This analysis was done with MDAnalysis's *distances* module.

To quantify the alignment of 2,4-D⁻ molecules on biochar surfaces, we calculated the angle between (i) aromatic ring and (ii) carboxylate group of 2,4-D⁻ and the biochar surface, defined by three closest biochar atoms that belong to the same molecular building block. This analysis was performed using MDAnalysis's *mdamath* module.

Furthermore, we calculated the Radial Distribution Function (RDF) between the cation and biochar surface, 2,4-D⁻ functional groups and water molecules. For this, we used MDAnalysis's *rdf* module.

## 2.5 Visualisation and Rendering

Snapshots of the simulation box were rendered with VMD v.1.9.4,[29] unless otherwise specified, carbon atoms are shown in cyan, oxygen atoms in red, chlorine atoms in green, hydrogen atoms in white, and sodium cation in blue. The periodic simulation box is shown in blue, water is always present but may not be shown for clarity. All plots were produced with Python 3.9 using the Python library Matplotlib.[30]

# 2 Results and Discussions

### 3.1 Characterisation of Biochar Molecular Models

The most essential step for a high-quality predictive simulation lies in the models used. To this end, in this work, we profit from the woody biochar models carefully developed by Wood *et al.*.[16] These models have been matched to a collection of target experimental data,[16] both chemical (H/C, O/C, aromatic domain size, aromaticity index, and functional groups) and physicochemical (true density) and morphological (HRTEM) descriptors. This approach ensures that the models are truly representative of real biochar materials produced from softwood feedstock at the highest treatment temperatures (HTT) of 400, 600 and 800 ºC – a typical temperature range used in biochar production. The properties of the three biochar models, BC400, BC600 and BC800, respectively, and the experimental property ranges used in the material development are given in **Table 1**.

*Table 1*. *Target physicochemical property (H/C and O/C molar ratio, aromaticity index and true density) ranges for biochar materials produced from softwood feedstock at the highest treatment temperature (HTT) of 400, 600 and 800 ºC and the corresponding properties of their representative models – BC400, BC600 and BC800. Confidence intervals for the experimental values are given in square brackets; standard deviations are given for the model's physicochemical descriptors (true density), based on three independent model replicas. Models are developed in Wood et al.[16] The experimental data collected and fitted in Wood et al.[19]*

| Experiment/Model | HTT (ºC) | H/C | O/C | Aromaticity index (%) | True density (kg m$^{-3}$) |
|---|---|---|---|---|---|
| Experiment | 400 | 0.65 [0.49-0.82] | 0.21 [0.15-0.29] | 75 [70-81] | 1430 [1380-1490] |
| Model BC400 | | 0.63 | 0.19 | 78 | 1402 ± 1 |
| Experiment | 600 | 0.23 [0.08-0.36] | 0.07 (0.02-0.15) | 96 [92-99] | 1540 [1490-1600] |
| Model BC600 | | 0.24 | 0.07 | 98 | 1581 ± 8 |
| Experiment | 800 | 0.12 [0.00-0.24] | 0.05 [0.00-0.12] | 99 [96-100] | 1850 [1800-1900] |
| Model BC800 | | 0.09 | 0.04 | 100 | 1878 ± 3 |

To be able to use these models for adsorption studies, the models must have an exposed surface, which was created in *xy*-plane by expanding a simulation box in the *z*-direction (**Figure S1.1**). These exposed surfaces feature a slightly different composition from the bulk material, as presented in **Table 2**. Exposed surfaces of lower temperature biochars (BC400 and BC600) have a lower H/C ratio than the bulk, suggesting that the exposed surfaces feature a higher proportion of aromatic domains. Yet, the O/C molar ratio is slightly higher on the surface for all models, indicating higher relative presence of functional groups at the surface. These functional groups vary with HTT – BC400 contains the largest quantity and variety of functional groups, whilst the higher temperature biochar models (BC600 and BC800) are increasingly more aromatic and, therefore, contain fewer and more stabilised functional groups (**Table 2**).

The surface of biochar materials is rough and cannot be fairly quantified by the *xy*-cross-section of the simulation box. Therefore, we calculate solvent accessible surface areas (SASAs) for each biochar system. Nevertheless, this value still remains dependent on the simulation system size and cannot be used directly for comparison, and so we also report normalised SASA (*n*SASA) per cross-sectional surface area (**Table 2**). *n*SASA of a completely flat surface would have a value of 1. Here, all materials show a degree of surface roughness (*n*SASA > 1) that increases with the HHT. Experimentally, the same trends are observed for the BET specific surface area (SSA) measurements.[19] While the models used in this work are not explicitly matched to represent intricate structures of microporous biochars – which is done in Ngambia *et al.*[31] – the increase in the surface area is also a measurement associated with microporosity, and is frequently used to characterise biochar materials.

*Table 2* Surface properties of BC400, BC600 and BC800 models: H/C molar ratio, O/C molar ratio, surface-exposed functional groups, cross-section surface area in xy-plane, solvent accessible surface area (SASA) calculated for $N_2$ with a probe radius 0.18 nm, and normalised SASA (nSASA) that demonstrates roughness of the surface (nSASA=1 indicates fully planar surface). H/C and O/C molar ratios in round brackets provide the bulk model's properties for comparison.

| Model | H/C | O/C | Functional groups | Cross-section area (nm$^2$) | SASA N$_2$ (nm$^2$) | *n*SASA (nm$^2$ nm$^{-2}$) |
|---|---|---|---|---|---|---|
| BC400-surf | 0.47 | 0.20 | carboxyl phenol carbonyl ester ether | 42.3 | 158.8 | 1.88 |
| BC600-surf | 0.19 | 0.08 | carbonyl ether phenol | 42.7 | 235.8 | 2.76 |
| BC800-surf | 0.15 | 0.07 | quinone pyrone | 105.5 | 798.3 | 3.78 |

## 3.2 2,4-D⁻ Uptake by Biochars

We begin by visually examining the simulation systems, with an example of equilibrated systems, given in **Figure 2** – a biochar material is submerged in a solution of 2,4-D⁻ molecules in water, charge balanced by Na⁺; a large proportion of 2,4-D⁻ is adsorbed onto biochar surface. This is a consistent observation for all of the starting solution concentrations. To quantify this further, we perform linear density analysis (**Figure S3.2**), which shows the distribution of species along the $z$-axis of the simulation box, i.e., perpendicular to the biochar surface. In all systems, biochar density fluctuated closely to its true density values (BC400 at ~1400 kg m$^{-3}$, BC600 at ~1580 kg m$^{-3}$ and BC800 at ~1880 kg m$^{-3}$), showing that even though the biochar models are made from individual building blocks, the material is stable. In all systems and concentrations, there is a visible increase of 2,4-D⁻ and Na⁺ density closer to the surface, which corresponds to adsorption. To quantify and compare the adsorption of 2,4-D⁻ by biochars, we first present the adsorption data in the form of isotherms and then discuss the adsorption mechanism.

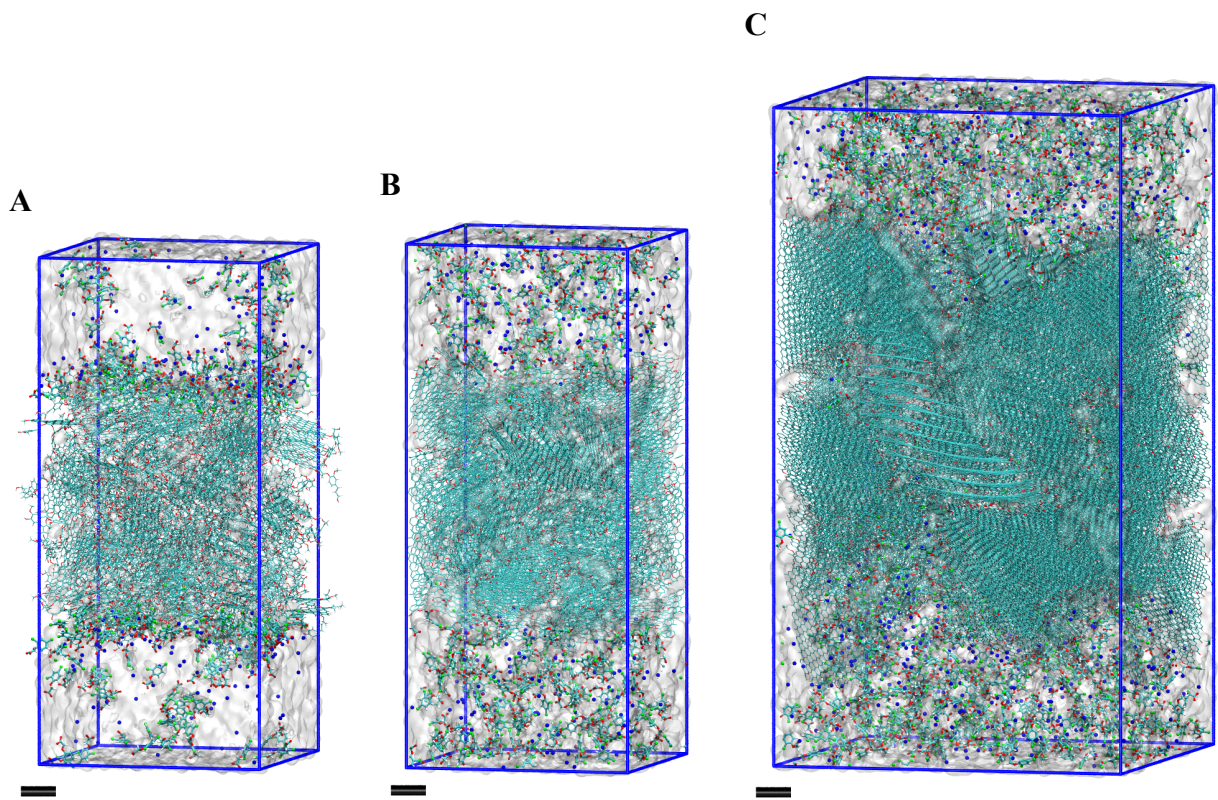

**Figure 2.** *Renderings of equilibrated A) BC400, B) BC600 and C) BC800 systems in 1.75 M 2,4-D solution. Biochar molecular units are shown in liquorice representations; 2,4-D⁻ molecules – CPK; Na⁺ ions – blue van der Waals spheres. Water is shown as transparent surface. Periodic simulation box is in blue. Scale bars are 1 nm.*

**Figure 3** presents three isotherm representations: adsorbed percentage vs starting concentration (**Figure 3A**); adsorbed amount per SASA of biochar vs equilibrium solution concentration (**Figure 3B**), and adsorbed amount per weight of biochar vs equilibrium solution concentration, which is convenient for comparison with experimental works (**Figure 3C**). It must be noted that while we can calculate the weight of a material in the simulation, this measurement is not informative, since when setting up the model we can vary, as desired, the ratio of bulk material (along *z*-direction) to exposed surface area (*xy*-pane). Only the surface area is important to the interfacial calculations, and the amount of bulk is chosen to be big enough to shield any long-range non-bonded interactions (i.e., it needs to be at least twice the vdW cut-off) but also not too large to balance resource use for simulation of this already many-atom system. Therefore, rather than calculating the mass of biochar in the system, we use $N_2$ SASA of the modelled biochar and convert it to the hypothetical weight of a material using a typical experimental $N_2$ BET-measured SSA for these materials (BC400 25 $m^2$ $g^{-1}$, BC600 100 $m^2$ $g^{-1}$, and BC800 150 $m^2$ $g^{-1}$).[19]

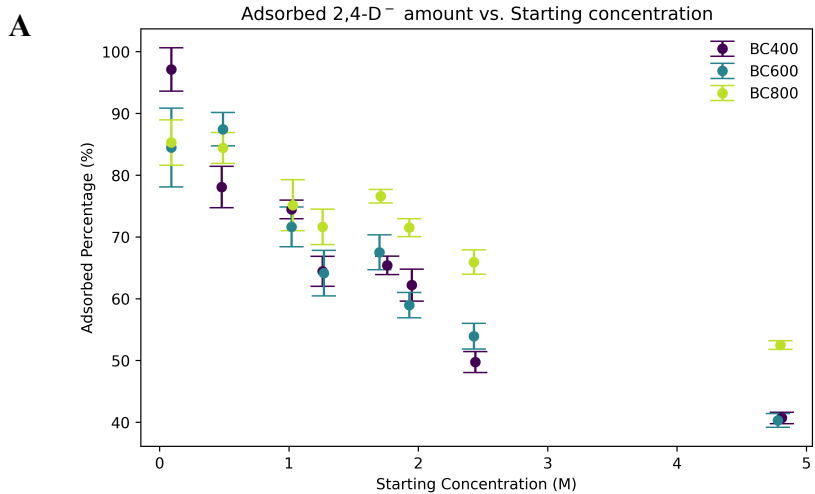
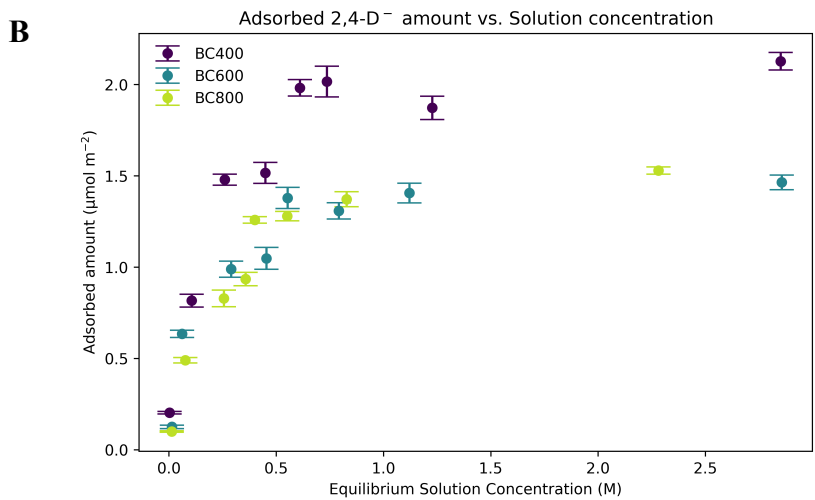
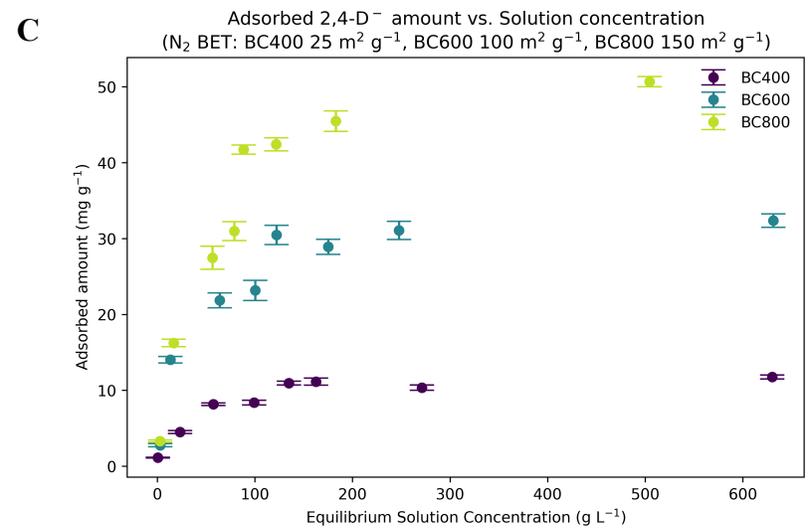

***Figure 3.** Adsorption isotherm of 2,4-D⁻ on biochar models: A) % adsorbed vs starting solution concentration; B) adsorbed amount per biochar surface area vs solution concentration at equilibrium; and C) adsorbed amount per gram of biochar vs solution concentration at equilibrium. Adsorbed amount is calculated as the number of 2,4-D⁻ molecules within 0 6 nm from the biochar surface; the mass of biochar is converted from models' $N_2$ SASA and using the experimental counterparts' $N_2$ BET surface area.*

When examining the percentage adsorption from a solution (**Figure 3A**), we can see that all biochar models follow the same trend – the lower the amount of 2,4-D⁻ in the solution, the higher is its removal, between 80% and 95% for the low concentration systems. This agrees with the experimental measures, where >80% removal is often reported for low concentrations.[32,33] These experimental studies should be interpreted carefully, as while a low concentration of 2,4-D⁻ used, the experiments are done with large volumes of pollutant solution to the amount of adsorbent, meaning that the number of 2,4-D⁻ molecules per biochar surface area is rather large. Therefore, on **Figure 3B**, we report adsorption per surface area of biochar. Interestingly, a very different picture is now seen. While higher temperature biochars (BC600 and BC800) trends are similar, with a maximum adsorption capacity of 1.5 µmol m⁻² (~0.33 mg m⁻²), the lower temperature BC400 shows a higher adsorption capacity per surface area of 2.1 µmol m⁻² (~0.47 mg m⁻²). This highlights that the adsorption is driven not only by surface area, but also by its chemical makeup. Yet, typically, in the experimental studies, the adsorption is reported as grams of pollutant per gram of adsorbent, as shown in **Figure 3C**, which would exaggerate the adsorption capacity for high-surface materials. Our data is in an excellent agreement with experimental studies,[33–36] where adsorption capacity generally ranges between 10-30 mg g⁻¹ and up to 50-70 mg g⁻¹ for high-performance materials; while a few chemically-activated biochars have been reported to have adsorption above 150 mg g⁻¹.[37]

### 3.3 Adsorption Mechanisms

Our simulations show that 2,4-D⁻ adsorption on woody biochar surfaces is not governed by a single dominant interaction, but by a combination of π-driven, polar and cation-mediated contributions. These interactions occur at distinct distances and orientations that correlate with specific functional groups on 2,4-D⁻ and with the evolving surface chemistry across BC400, BC600 and BC800 range. Below we discuss these mechanisms in terms of (i) aromatic interactions between 2,4-D⁻ and graphitic domains, (ii) interactions involving the carboxylate group and O-functionalities at biochar edges, and (iii) Na⁺-mediated bridging. We emphasise how these mechanisms depend on biochar production temperature and, hence, on aromaticity and surface functionality.

*3.3.1   Aromatic interactions: π – π and π – Cl contacts with biochar surface carbons*

The distances between the centre of mass (COM) of the 2,4-D⁻ aromatic ring and surface-exposed biochar carbons (and oxygens) show bimodality around ~3.6 Å and ~4.9 Å for all biochars and loadings (**Figures S3.3**, **S3.5**), which is not observed for hydrogen atoms (**Figure S3.4**). The distances to oxygen atoms are most prominent for higher temperature biochars (BC600 and BC800), where O-containing functional groups align with the plane of the aromatic carbons. Visualization of the adsorbed species at each measured distance (**Figure S3.6**), supported by angle distributions (**Figures S3.7**, **S3.8**) shows that these two populations correspond to distinct adsorption modes (illustrated in **Figure 4**):

- *Flat adsorption* (*Mode S1*) with the aromatic ring laying nearly parallel to the local biochar surface, at a distance of ~3.6 Å and with a tilt of <20°. This alignment is maximising π – π overlap between aromatic domains. Distances are consistent with stacked π – π contacts reported for graphitic systems and organic semiconductors.[38–40] While this mode is found on all three biochars, it is particularly favoured in regions with contiguous aromatic sheets.
- *Tilted adsorption* (*Mode S2*) with the aromatic ring near-perpendicular to the surface, at a distance of ~4.9 Å and with a tilt of ~40-80°, reminiscent of T-shaped π – π or edge-face contacts in conjugated systems.[38–40] This adsorption mode occurs both on flat graphitic patches and on rough, corrugated surface regions, including pores (**Figure 5A**) where steric constraints and local curvature disfavours fully parallel alignment of *Mode S1*.

In both modes, the chlorinated positions of 2,4-D⁻ remain equally close to the biochar surface. Probability distributions for Cl distances to surface C, H and O show broad single peaks near 3.4 Å, 2.8 Å and 3.2 Å, respectively (**Figures S3.9 – S3.11**), and angle distributions (**Figure S3.13**) confirm that 2,4-D⁻ orientations compatible with these contacts coincide with both *Modes S1* and *S2* (visualisation in **Figure S3.12**). These geometries are consistent with halogen – π (π – Cl) and halogen – O interactions superimposed on π – π binding, as reported for other chlorinated aromatics.[38] Our data, therefore, suggest that the chlorines do not define a separate adsorption mode, but rather stabilise existing π – π binding and edge contacts, likely contributing to the orientation of the adsorbed molecule on graphitic and functionalised sites.

Together, these observations support a previously identified mechanism, where π – π interactions between aromatic domains of 2,4-D⁻ and biochar graphitic domains are the necessary baseline mechanism across all HTT biochar ranges. Higher temperature produced BC600 and BC800, feature higher aromaticity and larger graphitic domains (**Table 1**) than BC400, and naturally provide more extended patches for π – π binding. Thus, higher HTT biochars show pronounced aromaticity-driven adsorption. However, the higher per-area adsorption capacity of BC400 (**Figure 3B**) indicates that π – π interactions alone do not explain overall uptake, and so the additional polar and cation-mediated interactions must noticeably contribute for the lower-temperature biochars with higher functionalised surfaces.

### 3.3.2 *Polar adsorption via the carboxylate group and edge functionalities*

The deprotonated carboxylate group of 2,4-D⁻ plays a central role in anchoring the molecule at polar sites on the biochar surface. Distances between the carboxylate COM and surface carbons show three main regimes at approximately 4.2 Å, 5.5 Å and 7.7 Å (**Figure S3.14**). Visualisation and angle analysis (**Figures S3.17 – S3.20**) reveal that these regimes correspond to three distinct adsorption geometries:

- *Fully parallel* (**Figure 4A**) adsorption with carboxylate groups within 4.9 Å and alike aromatic ring, is near-parallel to the surface (C–COO⁻ at < 20° to the surface). This corresponds to *Mode S1* adsorption, where π – π contacts are further reinforced by the carboxylate interaction to surface heteroatoms (O, H) and charge-balanced by Na⁺. These configurations are more common in, but not exclusive to, low-to-medium temperature biochars, where higher O/C ratios and more diverse O-functional groups are present.
- *Parallel aromatic ring only* (**Figure 4B**) adsorption, where the with aromatic domain remains parallel to the surface in *Mode S1*, while carboxylate groups are tilted away (C–COO⁻ to surface angle ~70°) with the carboxylate group found at 4.9 – 6.5 Å distance from the surface.
- *Tilted* (**Figure 4C**) adsorption in *Mode S2*, with the carboxylate protruding into the aqueous phase at a distance > 6.5 Å from the surface. Such states are likely to be weaker bound than either of the variations of *Mode S1*, and promote dynamic exchanges of molecules. These orientations are found at high concentrations, when the surface becomes saturated, as this alignment allows molecules to pack onto the surface.

Distance distributions between the carboxylate group and surface-exposed hydrogen and oxygen are broad and multimodal (**Figures S3.15**, **S3.16**), reflecting the structural heterogeneity of edge sites and nanopore interiors (**Figure S3.21**). Nevertheless, a clear peak at ~2.9 Å between –COO⁻ and surface H atoms on BC400, and to some extent on BC600, is consistent with H-bonding to surface –OH groups. Such –OH functionalities are abundant on lower temperature woody biochars and progressively lost at HTT >600 °C,[19,41] in line with our biochar model structures.[16] Visualisations show that some 2,4-D⁻ molecules are wedged into shallow nanopores or edge cavities, where the carboxylate can simultaneously interact with multiple sites (**Figure 5B** shows interaction within a shallow pore of BC400). These multipoint contacts likely provide an enthalpic contribution to adsorption.

Collectively, the above observations indicate that oxygen-rich, lower temperature produced biochars exploit a synergistic mechanism where the aromatic ring engages in π – π interactions with graphitic domains, while the deprotonated carboxylate group forms H-bonds with edge hydroxyl, carbonyl, and ether groups. This synergy explains why BC400 exhibits a higher adsorption capacity per unit area than BC600 and BC800 (**Figure 3B**), in which graphitised, oxygen-poor surfaces (BC800) provide little opportunity for direct H-bonding and adsorption relies predominantly on π – π interactions.

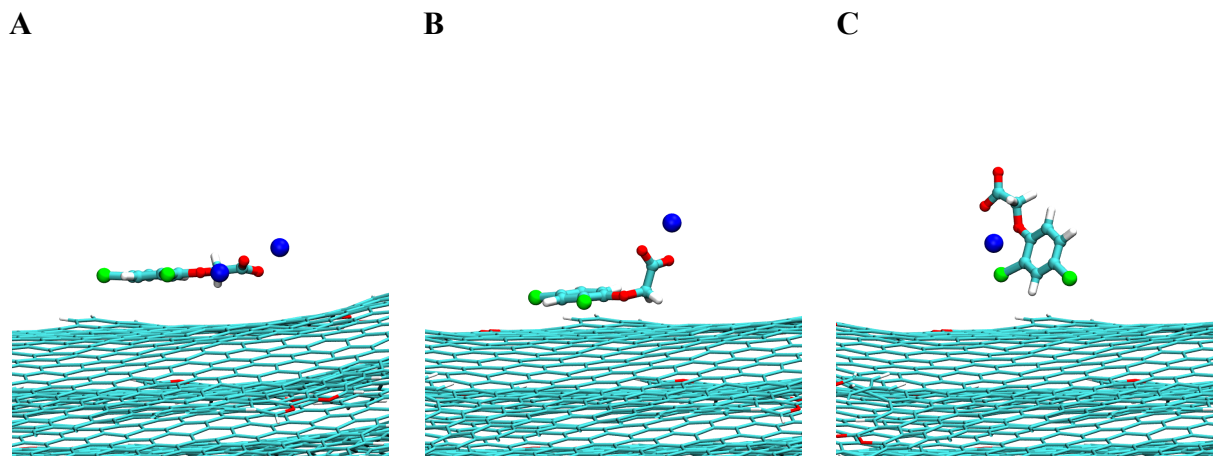

*Figure 4.* Representative visualisation of adsorption modes: A) completely parallel 2,4-D$^-$ alignment with the surface; B) 2,4-D$^-$ aromatic ring parallel to the surface and carboxylate group pointing away; and C) 2,4-D$^-$ is perpendicular to biochar surface, with chlorides pointing towards the surface.

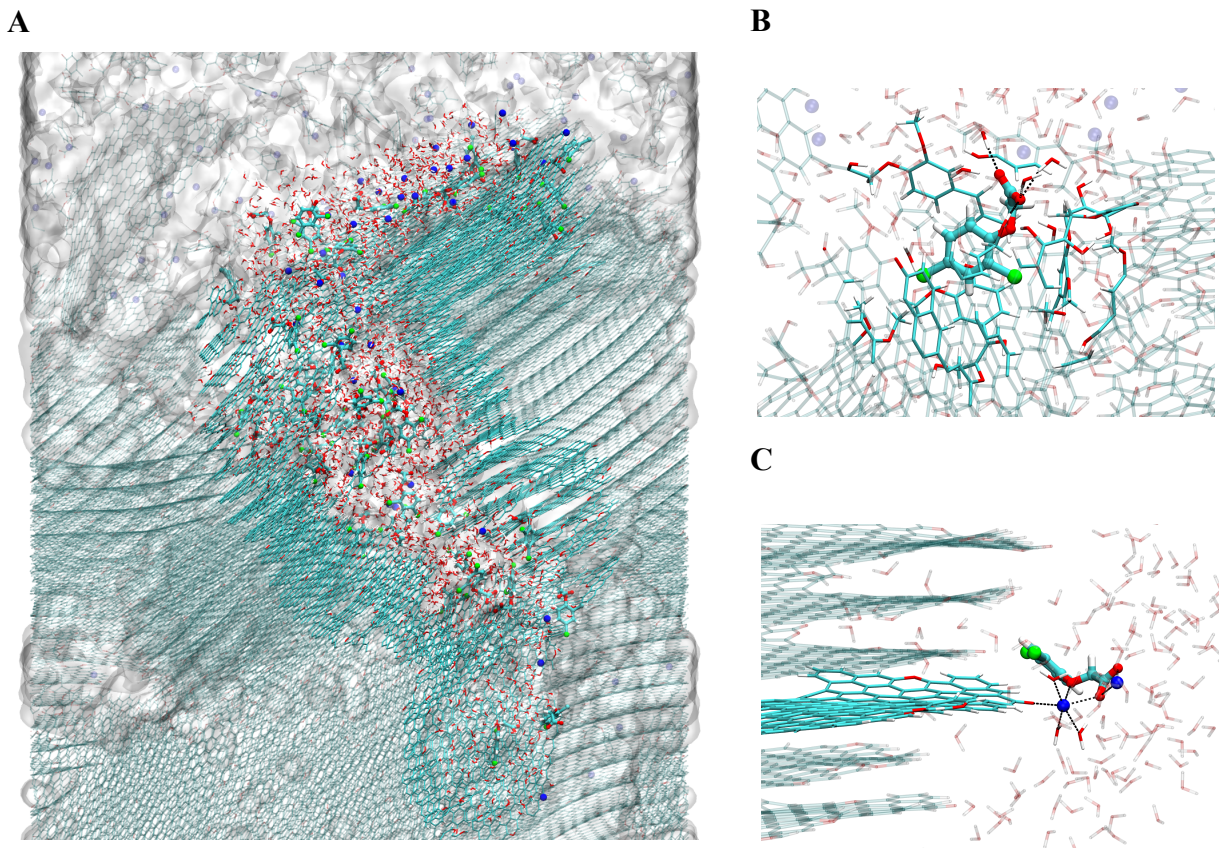

***Figure 5.*** *Illustration of A) 2,4-D⁻ diffusion and interaction within the pore channel of BC800; B) H-bond interaction between carboxylic group of 2,4-D⁻ and BC400 biochar nanopore; and C) Na⁺ mediated complex formation on the edge of a BC800 biochar-forming sheet. Molecules in the interaction site are shown in full opacity, while background is presented as transparent outlines/ surface water representation.*

### 3.3.3 *Cation bridging as Na$^+$ inner-sphere complex*

The third key component of the adsorption mechanism is the presence of charge-balancing cations in the system, here Na$^+$. The distances between Na$^+$ and biochar surface atoms and radial distribution functions from Na$^+$ to oxygens atoms of biochar, water and carboxylic group of 2,4-D$^-$ reveal a consistent strong peak at ~2.4 Å (**Figures S3.22 – 3.26**), characteristic of Na$^+$ – O complex distances.[42] In bulk solution, Na$^+$ is surrounded by six water molecules forming a first hydration shell. These water molecules can undergo a partial ligand exchange, forming an inner-sphere complex with either one or two carboxylate oxygens of 2,4-D$^-$ and/or biochar surface oxygen atoms (**Figures S3.27, S3.28**). **Figure 5C** illustrates a complex, where a single Na$^+$ cation simultaneously coordinates to both a carboxylate group, biochar surface oxygens, and waters. Distances between Na$^+$ and surface carbon atoms, defining the exposed surface, show double peaks at ~5.1 Å and ~7.4 Å, mirroring the two Na$^+$ – O distances at the surface (2.4 Å and 4.6 Å) and reflecting configurations where 2,4-D$^-$ is adsorbed via *Mode S1* or *S2*, respectively (**Figures S3.22, S3.23**). The above observations support the following mechanistic picture:

- Na$^+$ acts as a structure directing ion for 2,4-D$^-$ adsorption, replacing part of its hydration shell with surface and carboxylate oxygens.
- Cation bridging stabilises adsorption even on neutral or slightly negatively charged surfaces, where direct 2,4-D$^-$ anion-surface electrostatics would be unfavourable.
- On oxygen-rich surfaces of low-to-medium temperature biochars (BC400, BC600), Na$^+$ bridging occurs in addition to H-bonding, while on oxygen-poor highly aromatic surfaces of high-temperature produced biochars (BC800), it becomes a more critical polar contribution.

Given that the presence of ash makes biochars pH-rising compounds, that will acquire net negative charge above their point of zero charge (for woody biochars PZC is at pH≅7-7.5),[17] such cation mediated mechanisms are likely to be highly relevant under environmental conditions, especially in the presence of abundant Ca$^{2+}$ or Mg$^{2+}$. While our simulations explicitly include only Na$^+$, the same structural principles apply to multivalent cations, which would be expected to form even stronger bridging complexes with carboxylates and surface oxygen groups.

### 3.4 Biochar Design Implications

On the basis of discussed mechanistic insights, several recommendations emerge for the design and selection of biochars for the removal of 2,4-D and related chlorophenoxy anionic herbicides:

- ***Do not optimise for surface area alone: preserve oxygen-functionalities at moderate HTT.*** Our simulations show that low-temperature-produced biochars (here represented by BC400), which present moderate aromaticity while also featuring higher O/C and a larger density of oxygen-containing groups, exhibit higher adsorption per unit surface area than more graphitised materials produced at medium-to-high temperatures (here represented by BC600 and BC800). This enhancement arises from a cooperative interplay between $\pi - \pi$ interactions with graphitic domains and strong polar contacts, including H-bonding between the deprotonated 2,4-D$^-$ and surface hydroxyl, carbonyl, ester and ether functionalities  Therefore, for design of biochars aimed at anionic herbicides, pyrolysis temperatures in the approximate range of 350 °C to 500 °C are likely to be optimal, as those preserve substantial oxygen functionalities while still enabling some degree of aromatic structure. This perspective aligns with experimental trends reported for 2,4-D$^-$ and related species.[34,36]
- ***Exploit cation bridging at neutral to high pH, especially via divalent cations.*** 2,4-D is fully deprotonated in most environments (pKa=2.6), while biochar surfaces would be as they approach or exceed their PZC (pH≅7-7.5). In these cases, direct electrostatic attraction between 2,4-D$^-$ and the biochar surface weakens or becomes repulsive. For adsorption to occur in such regime, it must be stabilised by cation charge shielding or bridging. Common cations as Na$^+$, or even better if divalent Ca$^{2+}$ or Mg$^{2+}$, would be able to coordinate via inner-sphere complexation simultaneously with carboxylate oxygens of 2,4-D$^-$ and biochar surface oxygen functionalities. Therefore, for high-pH environments, biochars should feature accessible oxygen functionalities capable of coordinating cations and be deployed in environments with sufficient Ca$^{2+}$/Mg$^{2+}$, which could also come from biochar's own ash content. This mechanistic picture is consistent with the observed pH and ionic-strength dependencies of 2,4-D adsorption on biochars and cationic clays.[17,33,35]
- ***Utilise high-temperature, highly aromatic biochars when $\pi - \pi$ interactions dominate and cation availability is high.*** Biochars produced at higher temperatures (>600 °C) increasingly feature large graphitic domains, offering ample sites for $\pi - \pi$ and $\pi - $Cl interactions. However, it must be noticed that if comparing adsorption per surface area, their capacity is lesser than lower temperature biochars. Still, these high-temperature produced biochars offer significantly larger surface areas per unit of weight, which may compensate for loss of adsorption due to lower density of oxygen functionalities. Importantly, the higher temperature biochars (>800 °C) offer long-term structural persistence, important for applications where carbon sequestration is a priority. In such cases, presence of divalent cations can promote bridging, compensating for the reduced H-bonding capacity. In this context, post-treatments that mildly re-oxidise the surface (e.g., controlled air/ozone/steam activation) to reintroduce oxygen functionalities could further enhance 2,4-D affinity.[15,34,36,41]

- *Anticipate trade-offs between capacity, selectivity and reversibility.* Adsorption configurations in which the deprotonated carboxylate of 2,4-D$^-$ sits close to and parallel with the surface, supported by $\pi - \pi$, H-bonding and cation bridging (*Mode S1*-like geometries), are expected to be strongly bound and less reversible. This is of advantage for permanent removal, but may hinder desorption or adsorbent regeneration under, for example, changing soil conditions. In contrast, perpendicular configurations dominated by $\pi -$ Cl interactions (*Mode S2*) would be more weakly bound, more mobile, and readily exchangeable, offering greater reversibility. Consequently, HTT and post-treatment strategies can be tuned to balance capacity and reversibility, depending on whether the application emphasises single-use remediation or cyclic adsorption-desorption. Integrating these considerations with site-specific environmental conditions (pH, ionic composition, competing organics) will enable rational, application-driven design of biochars for anionic pollutant removal.

# 4 Conclusions

Through atomistic molecular dynamics simulations of experimentally constrained woody biochars, we demonstrate that 2,4-D adsorption is controlled by a combination of π-driven, polar and cation-mediated interactions, rather than by surface area alone. Across softwood-derived biochars produced between 400 and 800 °C, the aromatic ring of 2,4-D$^-$ adsorbs via parallel and perpendicular $\pi - \pi$ interactions with biochar's graphitic domains. This is further stabilised by proximity of the chlorinated positions to the surface. For the low-temperature produced O-rich biochars, these π-interactions are complemented by H-bonding and other polar contacts between the deprotonated carboxylate group and surface –OH and O-functionalities, leading to higher adsorption capacity per unit surface area than on more graphitised O-poor materials produced at higher temperatures.

In all systems, hydrated Na$^+$ forms inner-sphere complexes with 2,4-D$^-$ carboxylate oxygens and surface oxygen atoms, providing a clear molecular picture of cation bridging. This mechanism is expected to be even more important in the presence of environmental divalent Ca$^{2+}$ and Mg$^{2+}$, particularly at neutral to high pH where both biochar surfaces and 2,4-D$^-$ are negatively charged. Together, these results explain why experimental 2,4-D uptake does not correlate simply with specific surface area or bulk composition and highlight the need to consider surface functionality and solution chemistry when developing biochars for such applications.

More broadly, this work demonstrates that realistic, condensed-phase biochar molecular models can be used to both reproduce experimental isotherms and inspect the underlying adsorption mechanisms on complex carbonaceous materials. The same modelling framework can be used to guide the development of targeted biochars for removal of anionic herbicides and other contaminants by simultaneously tuning aromaticity, surface O-functionalities and interaction with environmental cations.

**Supporting Information**

Supporting information provided as "SI_24D_biochar.pdf". The data associated with this work, including structure files, topology files, force field files and MD input parameter files is freely available on GitHub: https://github.com/Erastova-group/24D_biochar, DOI: 10.5281/zenodo.17990328.

**Author Contributions**

R.W. carried out simulations, collected and analysed the data, and drafted the manuscript. V.E. acquired funding, supervised the project, analysed data, written the final version of the manuscript. All authors have jointly conceived the presented idea and contributed to the manuscript preparation and its review.

**Funding Sources**

NERC Doctoral Training Partnership grant (NE/S007407/1). Cirrus EPSRC access (EP/P020267/1).

**Acknowledgment**

R.W. would like to thank the NERC Doctoral Training Partnership grant (NE/S007407/1) for the funding of her MRes project "Molecular Modeling for Design of Biochar Materials." All simulations were performed on the Cirrus UK National Tier-2 HPC Service at EPCC (http://www.cirrus.ac.uk) funded by the University of Edinburgh and EPSRC (EP/P020267/1).

Supplementary Information

for

# Unravelling 2,4-D – biochar interactions by molecular dynamics: adsorption modes and surface functionalities

*Rosie Wood,[1] Ondřej Mašek,[2] and Valentina Erastova[1]\**

[1] School of Chemistry, University of Edinburgh, Joseph Black Building, David Brewster Road, King's Buildings, Edinburgh EH9 3FJ, UK

[2] UK Biochar Research Centre, School of GeoSciences, University of Edinburgh, Crew Building, Alexander Crum Brown Road, King's Buildings, Edinburgh EH9 3FF, UK

\* valentina.erastova@ed.ac.uk

## S1. SYSTEM SET-UP

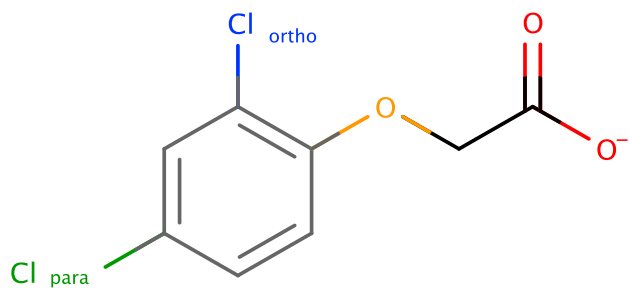

*Figure S1.0*. Structure of deprotonated 2,4-dichlorophenoxyacetic acid (2,4-D$^-$). The carboxylate oxygens are shown in red, the ether oxygen is shown in orange, the aromatic carbons are shown in grey, and the two chlorines (para and ortho) are shown in green and blue, respectively.

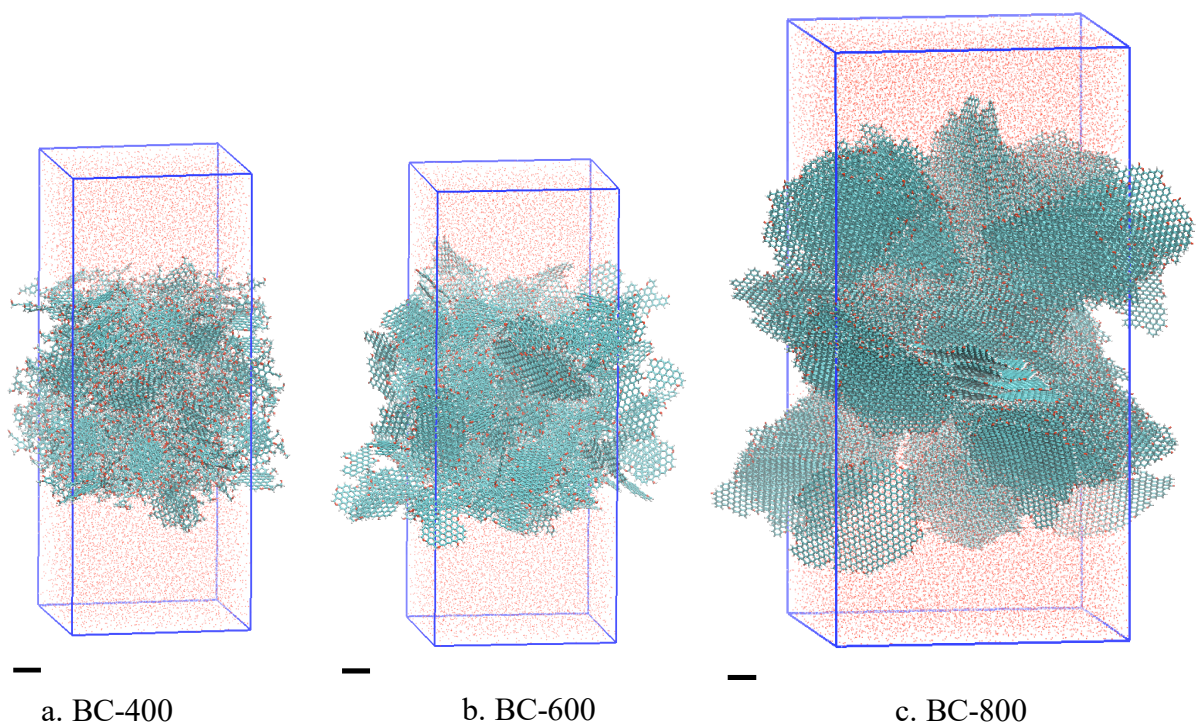

a. BC-400      b. BC-600      c. BC-800

*Figure S1.1*. Visualisation of a) BC400, b) BC600, and c) BC800 biochar systems with exposed surface. Atoms shown in cyan (C), red (O) and white (H). The periodic simulation box is given in blue. Scale bars are 1 nm.

*Table S1.1*. *System composition and simulation box sizes of B400, B600 and B800 biochars.*

| System ID | Number of atoms | | | $x \times y \times z$ dimensions, nm (incl. free space in $z$-dir) |
|---|---|---|---|---|
| | *C* | *H* | *O* | |
| BC400 | 14,560 | 9,240 | 2,800 | 6.50611 × 6.50611 × 15.15675 |
| BC600 | 19,040 | 4,620 | 1,400 | 6.53285 × 6.53285 × 15.02728 |
| BC800 | 87,920 | 7,840 | 3,080 | 10.26923 × 10.26923 × 18.71846 |

*Table S1.2*. *Molecular compositions of the B400, B600 and B800 systems with various concentrations of $Na^+$ 2,4-$D^-$.*

| System ID | $Na^+$ 2,4-$D^-$ molecules | Water molecules | Initial concentration (M) |
|---|---|---|---|
| BC400-C1 | 20 | 12286 | 0.09 |
| BC400-C2 | 100 | 11469 | 0.48 |
| BC400-C3 | 190 | 10349 | 1.02 |
| BC400-C4 | 225 | 9897 | 1.26 |
| BC400-C5 | 290 | 9133 | 1.76 |
| BC400-C6 | 310 | 8839 | 1.95 |
| BC400-C7 | 360 | 8183 | 2.44 |
| BC400-C8 | 500 | 5773 | 4.81 |

| System ID | $Na^+$ 2,4-$D^-$ molecules | Water molecules | Initial concentration (M) |
|---|---|---|---|
| BC600-C1 | 21 | 12670 | 0.09 |
| BC600-C2 | 103 | 11728 | 0.49 |
| BC600-C3 | 196 | 10630 | 1.02 |
| BC600-C4 | 232 | 10174 | 1.27 |
| BC600-C5 | 290 | 9454 | 1.70 |
| BC600-C6 | 315 | 9060 | 1.93 |
| BC600-C7 | 370 | 8439 | 2.43 |
| BC600-C8 | 516 | 5988 | 4.78 |

| System ID | $Na^+$ 2,4-$D^-$ molecules | Water molecules | Initial concentration (M) |
|---|---|---|---|
| BC800-C1 | 56 | 34251 | 0.09 |
| BC800-C2 | 279 | 31700 | 0.49 |
| BC800-C3 | 530 | 28554 | 1.03 |
| BC800-C4 | 627 | 27670 | 1.26 |
| BC800-C5 | 790 | 25679 | 1.71 |
| BC800-C6 | 860 | 24671 | 1.93 |
| BC800-C7 | 1000 | 22820 | 2.43 |
| BC800-C8 | 1400 | 16179 | 4.80 |

## S2. CONVERGENCE OF THE SYSTEMS

All of the systems have been accessed to ensure they have converged. This allowed us to establish what part of trajectory can be used for analysis. We have done the assessment using traditional RMSD (**Fig. S2.1**) as well as *DynDen*[1] (**Fig. S2.2, S2.3**) – our own tool to access the convergence of surface-exposed systems.

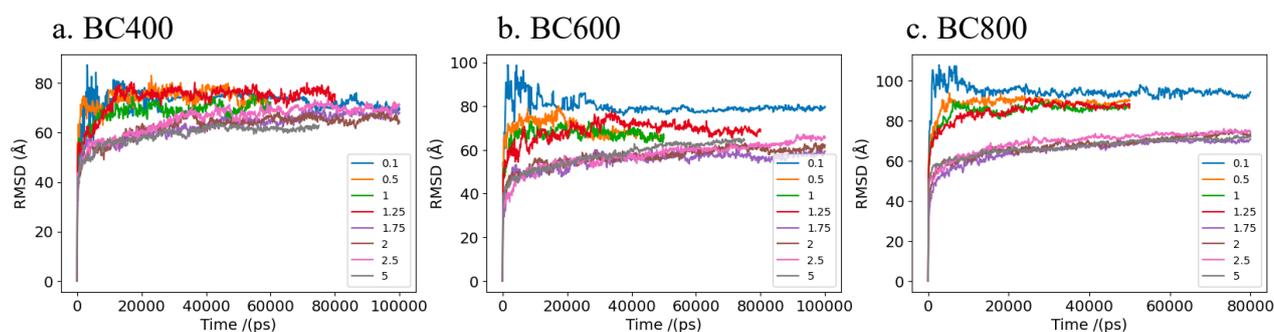

***Figure S2.1.*** *Root Mean Square Deviations (RMSDs) of 2,4-D$^−$ molecules across adsorption simulation trajectories in a) BC400, b) BC600, and c) BC800 systems. Legend shows 2,4-D$^−$ concentrations in mol L$^{-1}$.*

---

[1] Degiacomi, M. T., Tian, S., Greenwell, H. C. & Erastova, V. DynDen: Assessing convergence of molecular dynamics simulations of interfaces. *Comput Phys Commun* 269, (2021).

a.

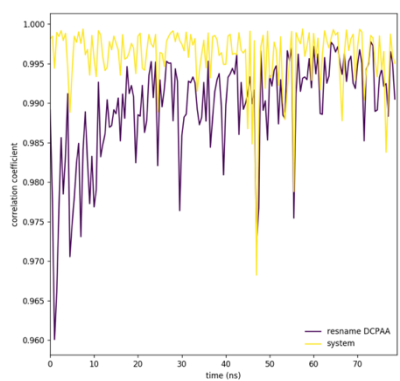

b.

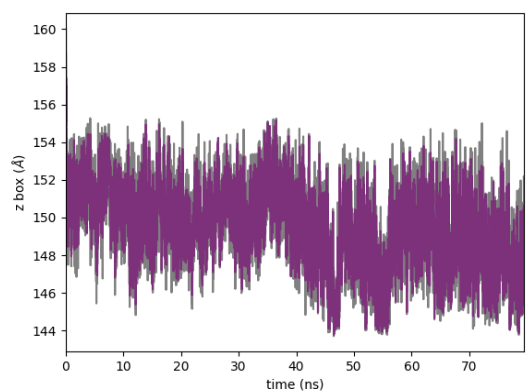

c.

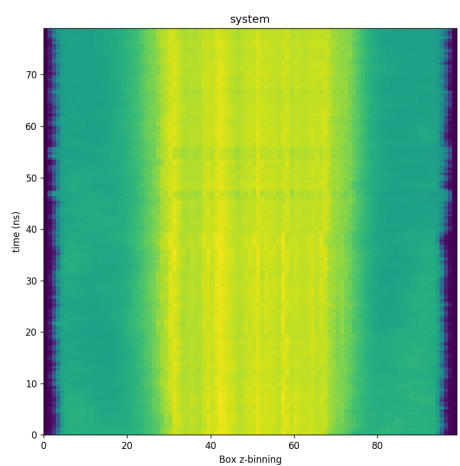

d.

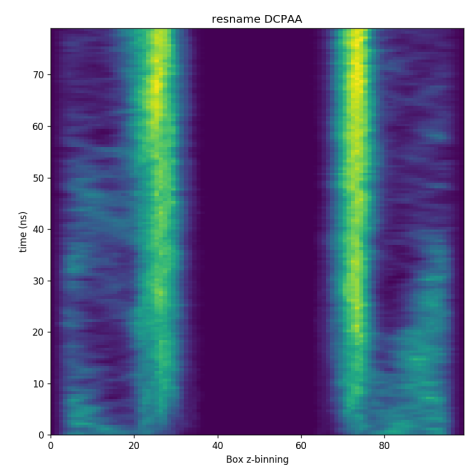

e.

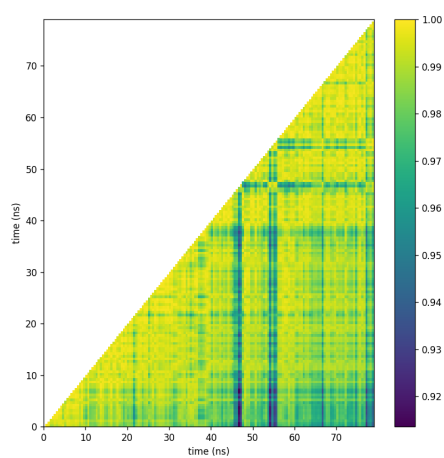

f.

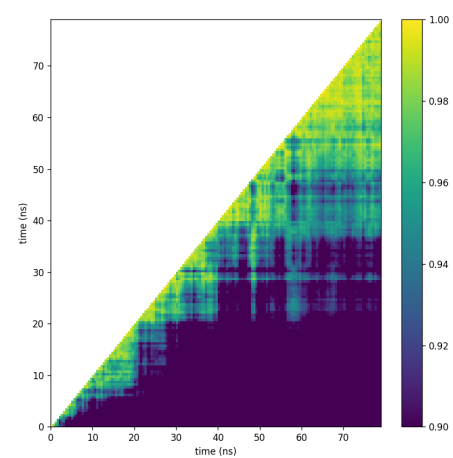

***Figure S2.2**. An example of full DynDen output for BC600 with 1.23 mol L$^{-1}$ 2,4-D$^-$. a) shows RMSD for whole system and 2,4-D$^-$; b) shows changes in the z-direction of simulation box over time; c) shows linear density evolution over time for the whole system, d) for 2,4-D$^-$ only; e) shows cross-correlation between frames for the whole system, and f) for 2,4-D$^-$ only. It can be seen from (f) that convergence is occurring after 35 ns for the most mobile 2,4-D$^-$ component.*

| System | 0.1 M 2,4-D⁻ | 5 M 2,4-D⁻ |
|---|---|---|
| BC-400 | 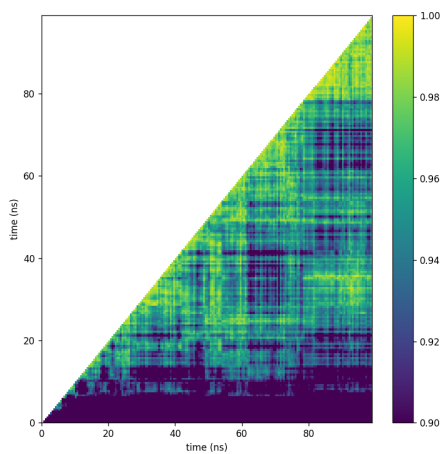 | 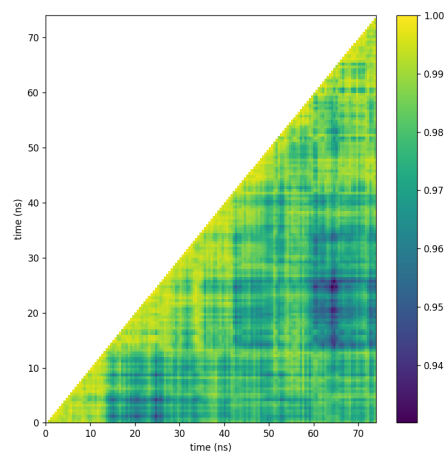 |
| BC-600 | 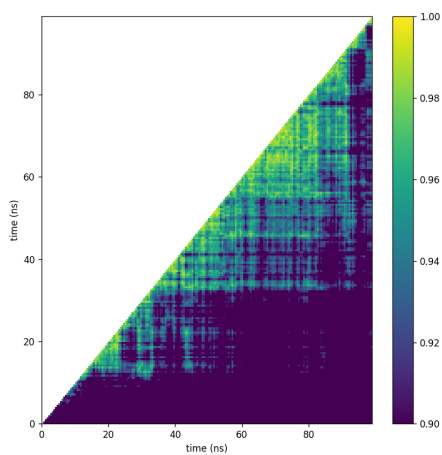 | 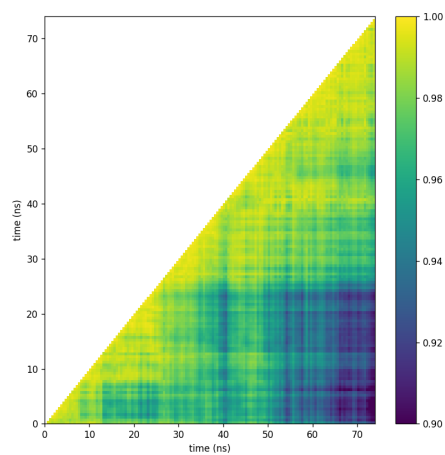 |
| BC-800 | 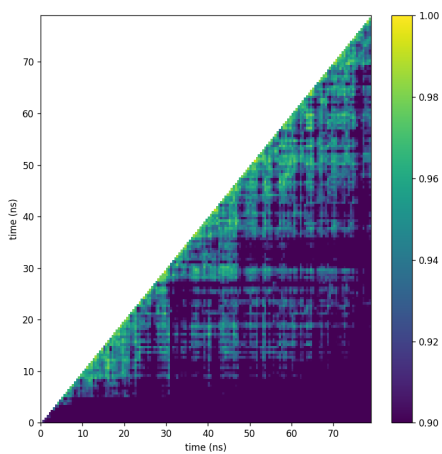 | 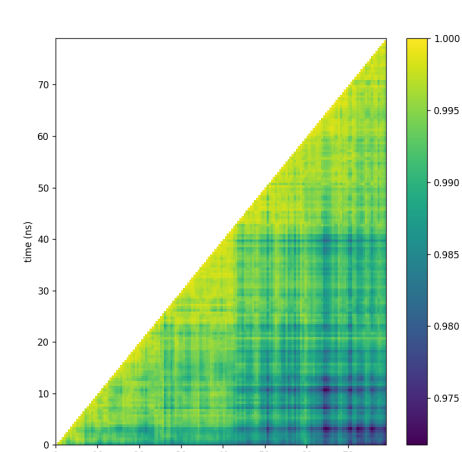 |

*Figure S2.3.* DynDen output for cross-correlation of most mobile component, i.e., 2,4-D⁻, for lowest (0.1 M, left column) and highest (5 M, right column) of BC400 (top row), BC600 (middle row) and BC800 (bottom row) systems at highest and lowest concentrations.

## S3. SIMULATION ANALYSIS

**Surface-exposed atoms**

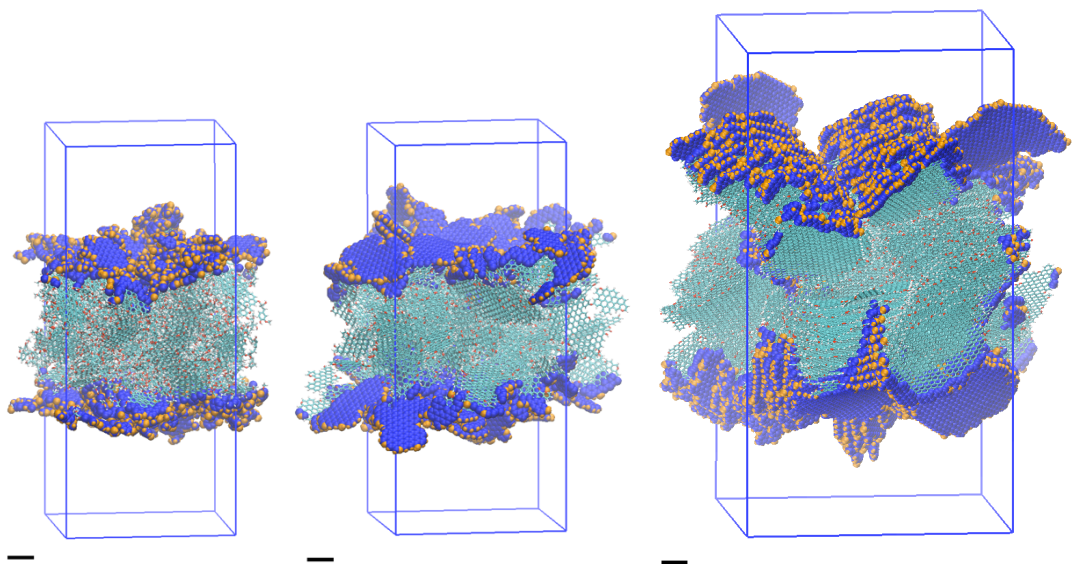

*Figure S3.1*. Visualisation of surface-exposed C atoms (blue) and surface-exposed H and O atoms (orange) of BC400, BC600 and BC800. Scale bars are 1 nm.

**Linear density profiles**

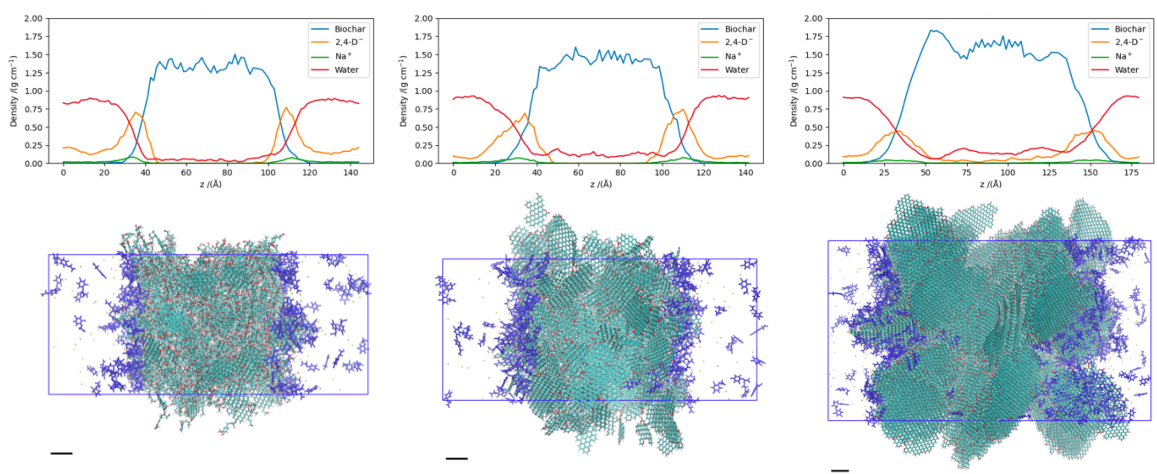

*Figure S3.2*. Example linear densities and visualisations of a) BC400-surf, b) BC600-surf and c) BC800-surf systems after equilibration containing a 1.75 M solution of 2,4-D, counterbalanced by $Na^+$. Biochar models are coloured in cyan (C), red (O), and white (H), 2,4-D$^-$ molecules are coloured in blue and $Na^+$ ions are coloured in orange. Water molecules not shown for clarity. Scale bars are 1 nm.

## S3.1 AROMATIC RING OF 2,4-D TO BIOCHAR SURFACE ATOMS: C, H, O

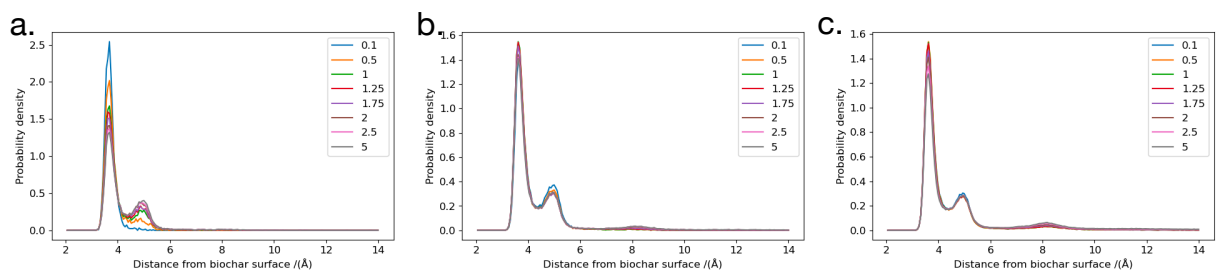

*Figure S3.3.* Probability density plots showing distances between centre of mass of 2,4-D⁻ aromatic rings and surface-exposed biochar C atoms after equilibration. Each plot shows systems containing a) BC400 b) BC600 and c) BC800. Starting 2,4-D⁻ concentration is shown in legend in units of mol L$^{-1}$.

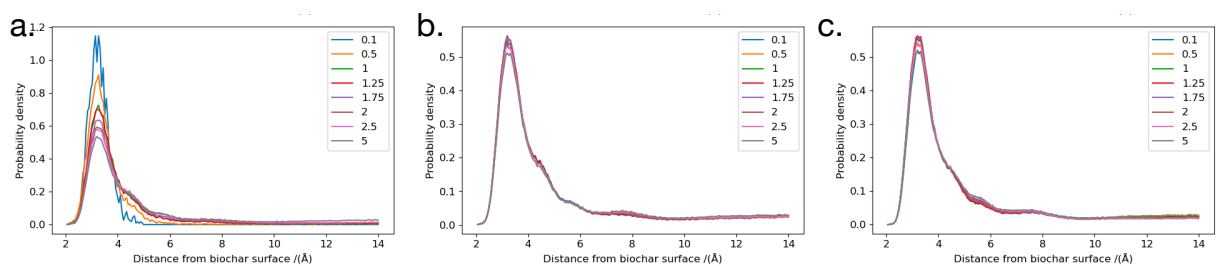

*Figure S3.4.* Probability density plots showing distances between the centre of mass of 2,4-D⁻ aromatic rings and biochar H atoms after equilibration. Each plot shows systems containing a) BC400-surf, b) BC600-surf and c) BC800-surf. Starting 2,4-D⁻ concentration is shown in legend in units of mol L$^{-1}$.

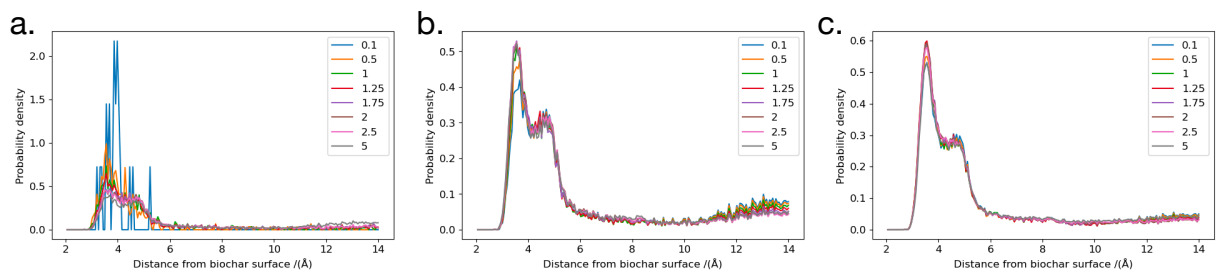

*Figure S3.5.* Probability density plots showing distances between the centre of mass of 2,4-D⁻ aromatic rings and biochar O atoms after equilibration. Each plot shows systems containing a) BC400-surf, b) BC600-surf and c) BC800-surf. Starting 2,4-D⁻ concentration is shown in legend in units of mol L$^{-1}$.

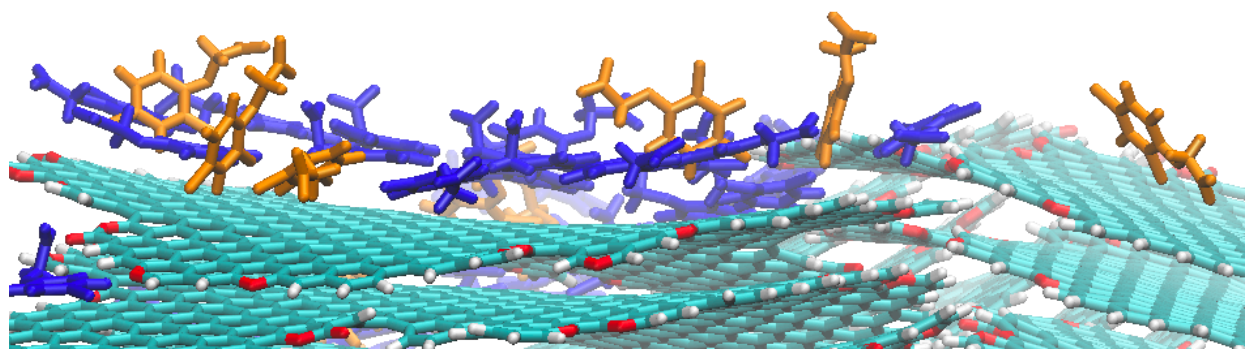

*Figure S3.6.* Representative visualisation of an equilibrated system. Biochar model is coloured in cyan (C), red (O) and white (H); 2,4-D⁻ molecules with their aromatic ring adsorbed to surface-exposed biochar C atoms at distances < 4.2 Å are coloured in blue (Mode S1) and 2,4-D⁻ molecules with their aromatic ring adsorbed to surface-exposed biochar C atoms at distances > 4.2 Å are coloured in orange (Mode S2). $Na^+$ ions and water molecules are not shown for clarity.

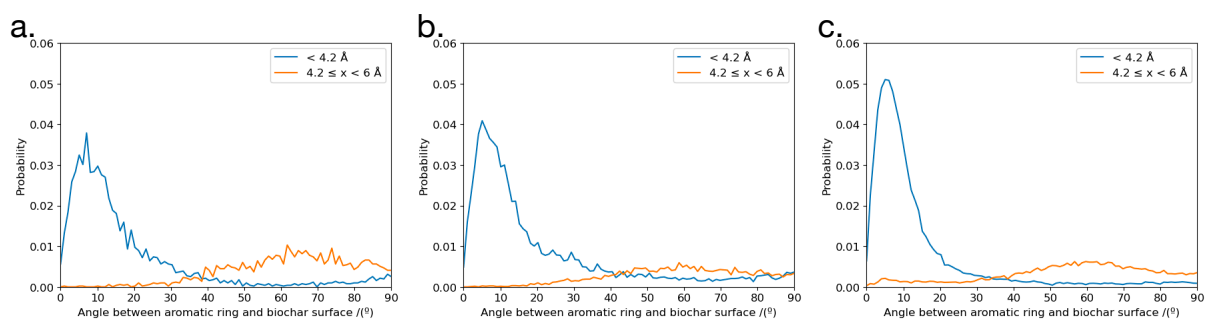

*Figure S3.7.* Example of variation in angle between 2,4-D⁻ aromatic ring and biochar surface for 2,4-D⁻ molecules adsorbed to surface-exposed C atoms. Plots show result for systems containing a) BC400-surf, b) BC600-surf and c) BC800-surf and a 1 mol L⁻¹ starting concentration of 2,4-D⁻. Where Mode S1 is found at the distance < 4.2 Å, and Mode S2 is found between 4.2 and 6 Å.

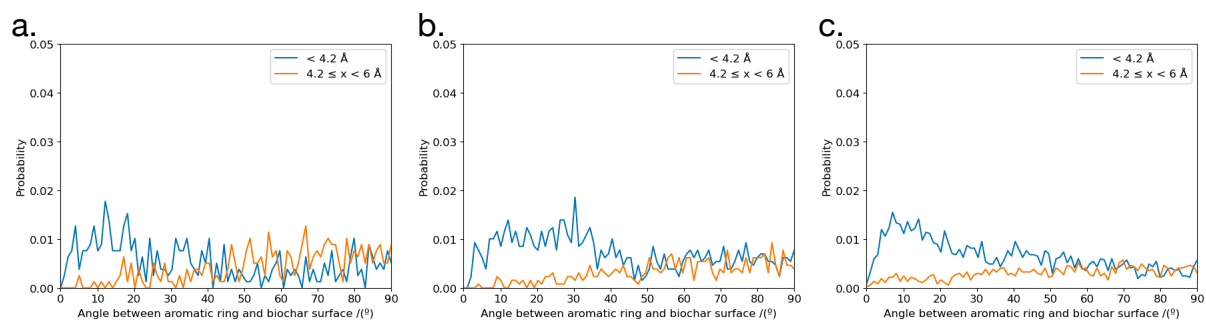

*Figure S3.8.* Example of variation in angle between 2,4-D⁻ aromatic ring and biochar surface for 2,4-D⁻ molecules adsorbed to surface-exposed O atoms. Plots show result for systems containing a) BC400-surf ,b) BC600-surf and c) BC800-surf and a 1 mol L⁻¹ starting concentration of 2,4-D⁻. Where Mode S1 is found at the distance < 4.2 Å, and Mode S2 is found between 4.2 and 6 Å.

## S3.2 CLORINE ATOMS OF 2,4-D TO BIOCHAR SURFACE ATOMS: C, H, O

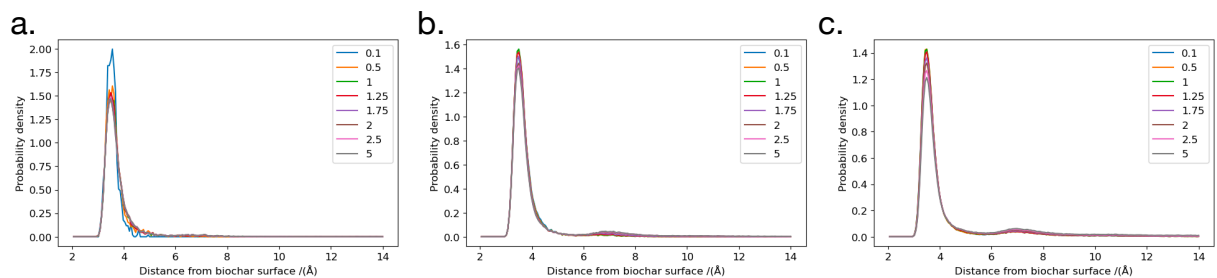

*Figure S3.9*. Probability density plots showing distances between Cl atoms and surface-exposed biochar C atoms after equilibration. Each plot shows systems containing a) BC400-surf b) BC600-surf and c) BC800-surf Starting 2,4-D⁻ concentration is shown in legend in units of mol L$^{-1}$.

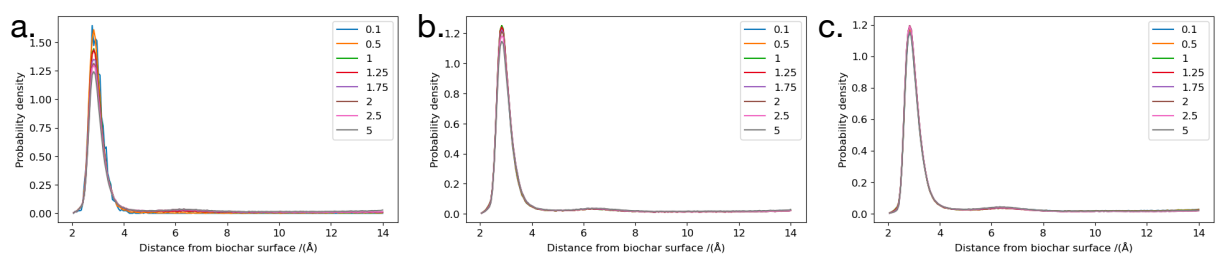

*Figure S3.10*. Probability density plots showing distances between the Cl atoms of 2,4-D⁻ and surface-exposed biochar H atoms after equilibration. Each plot shows systems containing a) BC400-surf, b) BC600-surf and c) BC800-surf. Starting 2,4-D⁻ concentration is shown in legend in units of mol L$^{-1}$.

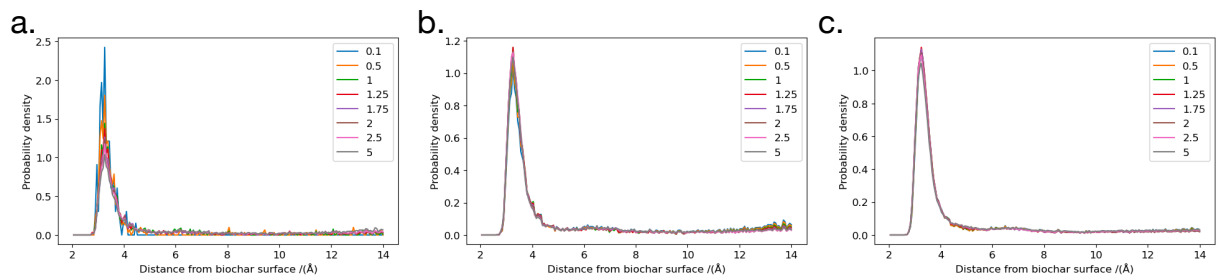

*Figure S3.11*. Probability density plots showing distances between the Cl atoms of 2,4-D⁻ and surface-exposed biochar O atoms after equilibration. Each plot shows systems containing a) BC400-surf b) BC600-surf and c) BC800-surf. Starting 2,4-D⁻ concentration is shown in legend in units of mol L$^{-1}$.

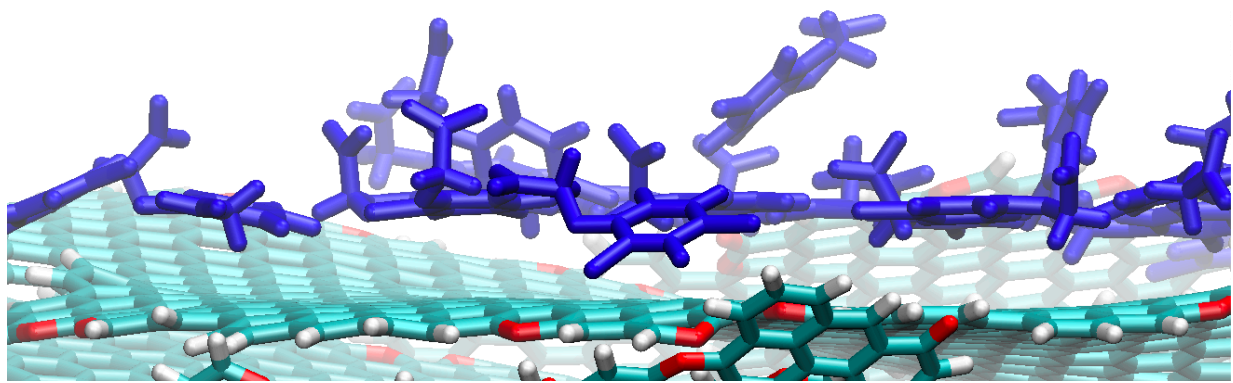

*Figure S3.12. Representative visualisation of an equilibrated system. Biochar model is coloured in cyan (C), red (O) and white (H), 2,4-D⁻ molecules with their $Cl_{para}$ atoms adsorbed to surface-exposed C atoms are coloured in blue. $Na^+$ ions and water molecules not shown for clarity.*

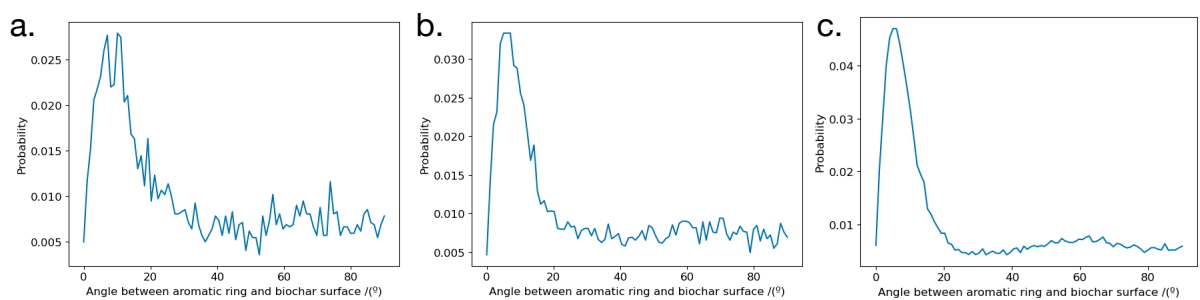

*Figure S3.13. Example of variation in angle between 2,4-D⁻ aromatic ring and biochar surface for 2,4-D⁻ molecules adsorbed with Cl groups at 3.0 – 4.0 Å to surface-exposed C atoms of biochar. Plots show results for systems containing a) BC400-surf, b) BC600-surf, and c) BC800-surf and a 1 mol $L^{-1}$ starting concentration of 2,4-D⁻.*

## S3.3 CARBOXYLATE GROUP OF 2,4-D TO BIOCHAR SURFACE ATOMS: C, H, O

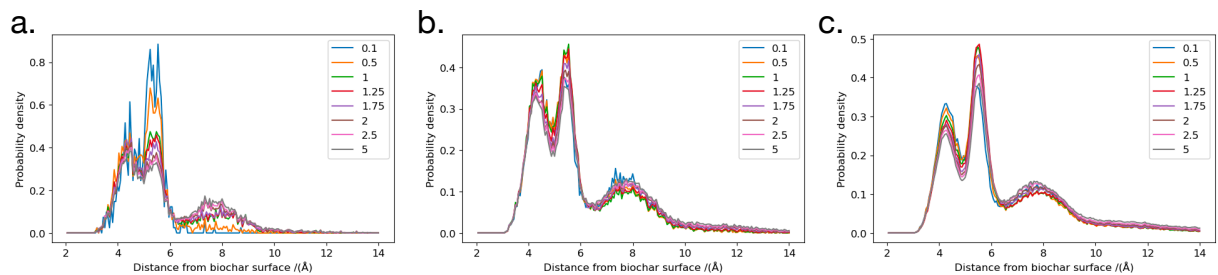

*Figure 3.14*. Probability density plots showing distances between the center of mass of carboxylate (COO$^-$) group and surface-exposed biochar C atoms after equilibration. Each plot shows systems containing a) BC400-surf, b) BC600-surf and c) BC800-surf. Starting 2,4-D$^-$ concentration is shown in legend.

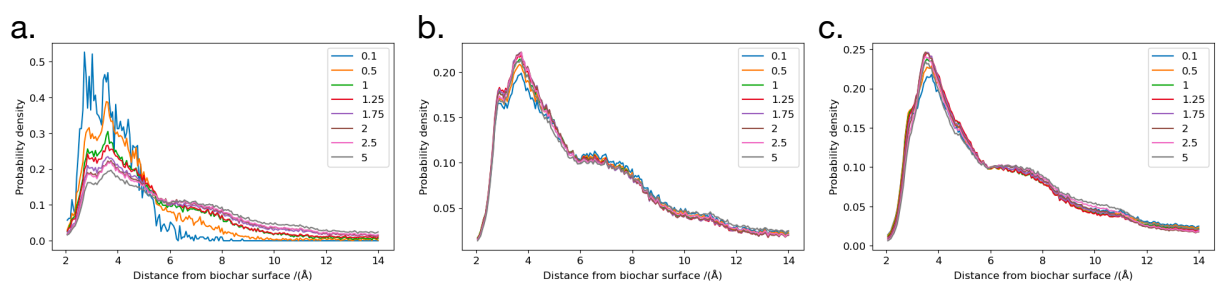

*Figure S3.15*. Distances between the centre of mass of 2,4-D$^-$ carboxylate (COO$^-$) and surface-exposed biochar H atoms after equilibration. Each plot shows systems containing a) BC400-surf b) BC600-surf and c) BC800-surf. Starting 2,4-D$^-$ concentration is shown in legend in units of mol L$^{-1}$.

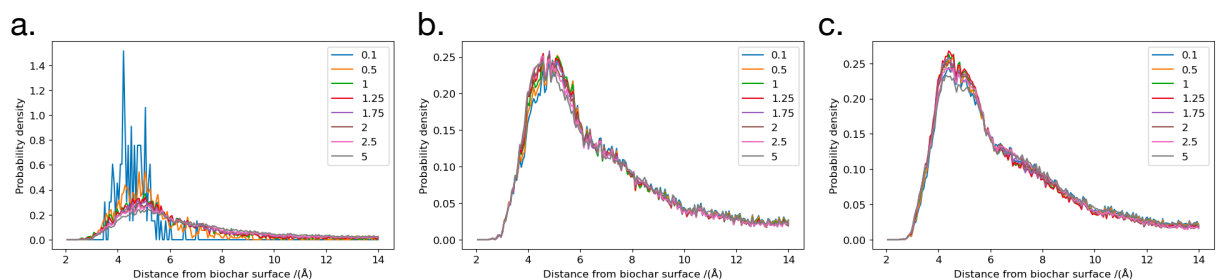

*Figure S3.16*. Distances between the centre of mass of 2,4-D$^-$ carboxylate (COO$^-$) and surface-exposed biochar O atoms after equilibration. Each plot shows systems containing a) BC400-surf, b) BC600-surf and c) BC800-surf. Starting 2,4-D$^-$ concentration is shown in legend in units of mol L$^{-1}$.

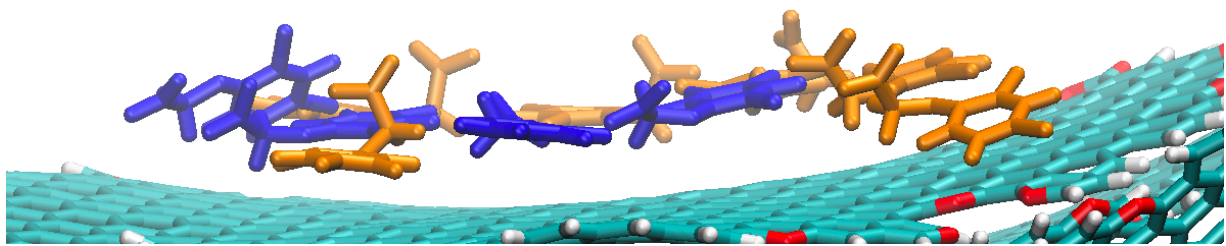

***Figure S3.17***. *Representative visualisation of an equilibrated system. Biochar model is coloured in cyan (C), red (O) and white (H); 2,4-D⁻ molecules with their carboxylate group adsorbed to surface sites at distances < 4.9 Å are coloured in blue and 2,4-D⁻ molecules with their carboxylate group adsorbed to surface sites at distances of 4.9 Å - 6.5 Å are coloured in orange. $Na^+$ ions, water molecules and 2,4-D⁻ molecules with their carboxylate group adsorbed to surface sites at distances of greater than 6.5 Å are not shown for clarity.*

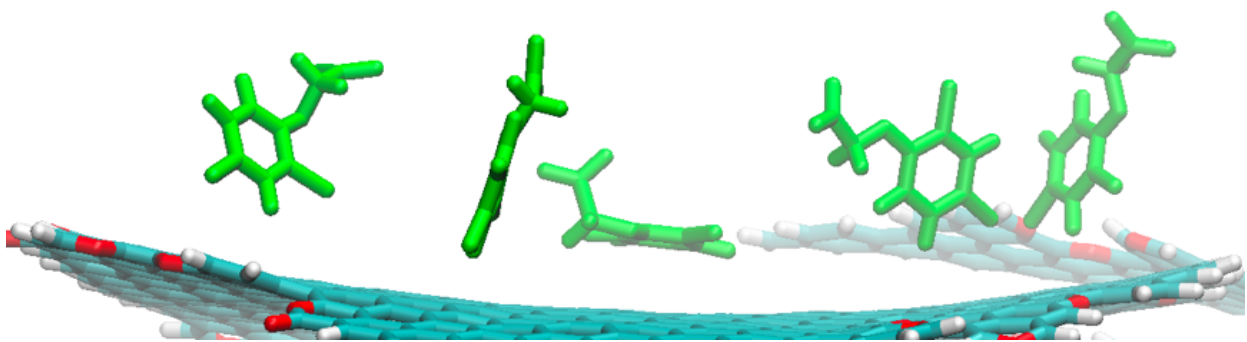

***Figure 3.18***. *Representative visualisation of an equilibrated system. Biochar model is coloured in cyan (C), red (O) and white (H); 2,4-D⁻ molecules with their carboxylate group adsorbed to surface sites at distances > 6.5 Å are coloured in green. $Na^+$ ions, water molecules and 2,4-D⁻ molecules with their carboxylate groups adsorbed to surface sites at distances < 6.5 Å are not shown for clarity.*

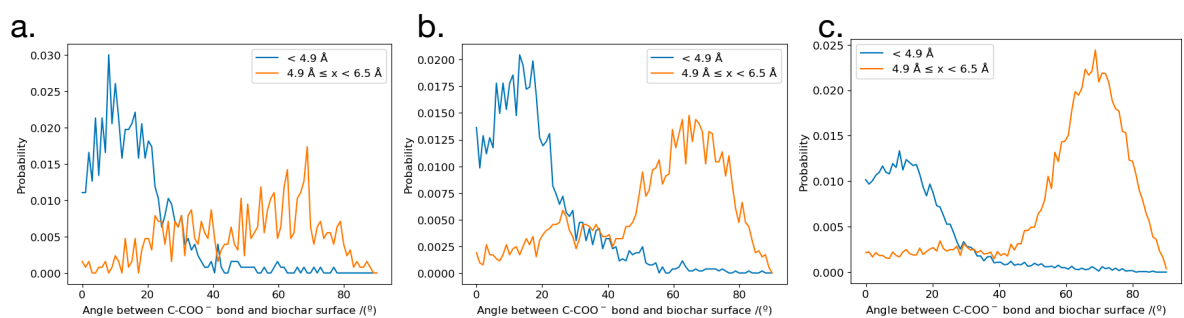

***Figure S3.19***. *Example of variation in angle between 2,4-D⁻ C-COO bond and biochar surface for 2,4-D⁻ molecules adsorbed to surface sites. Plots show result for systems containing a) BC400, b) BC600, and c) BC800 and a 1 mol $L^{-1}$ starting concentration of 2,4-D⁻.*

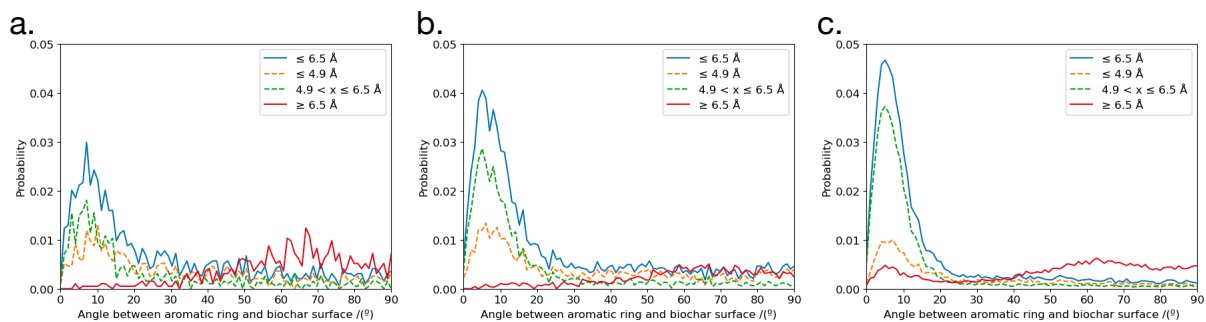

***Figure S3.20***. *Example of variation in angle between 2,4-D⁻ aromatic ring and biochar surface for 2,4-D⁻ molecules adsorbed to surface-exposed C atoms. Plots show result for systems containing a) BC400-surf, b) BC600-surf and c) BC800-surf and a 1 mol L$^{-1}$ starting concentration of 2,4-D⁻.*

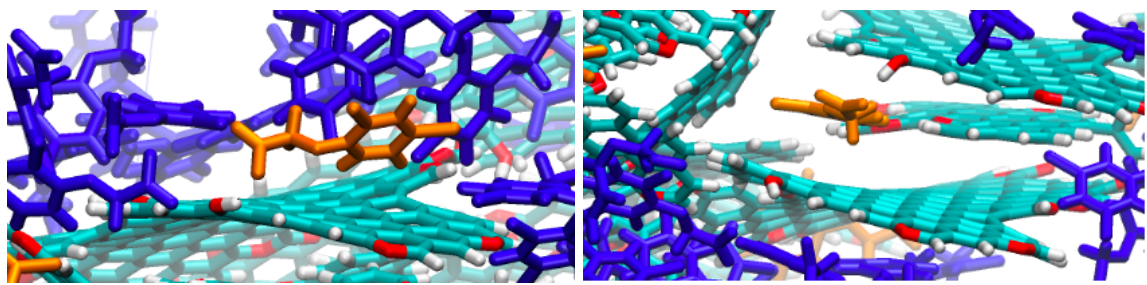

***Figure S3.21***. *Representative visualisation of an equilibrated system. Biochar model is coloured in cyan (C), red (O) and white (H), 2,4-D⁻ molecules with their carboxylate group adsorbed to biochar edge sites at any distance are coloured in blue and 2,4-D⁻ molecules with their carboxylate group adsorbed to biochar H atoms at distances of < 3.1 Å are coloured in orange. Na$^+$ ions and water molecules not shown for clarity.*

## S3.4 Interactions between 2,4-D and biochar mediated by Na$^+$ ions

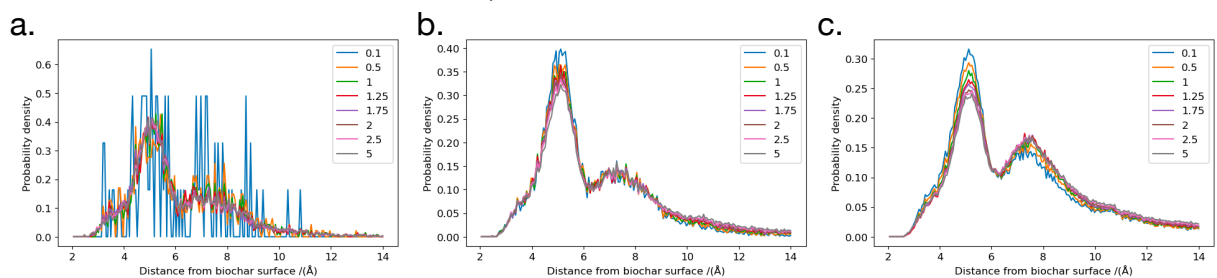

*Figure S3.22*. Probability density plots showing distances between Na$^+$ ions and surface-exposed biochar C atoms after equilibration. Each plot shows systems containing a) BC400-surf, b) BC600-surf and c) BC800-surf. Starting 2,4-D$^-$ concentration is shown in legend in units of mol L$^{-1}$.

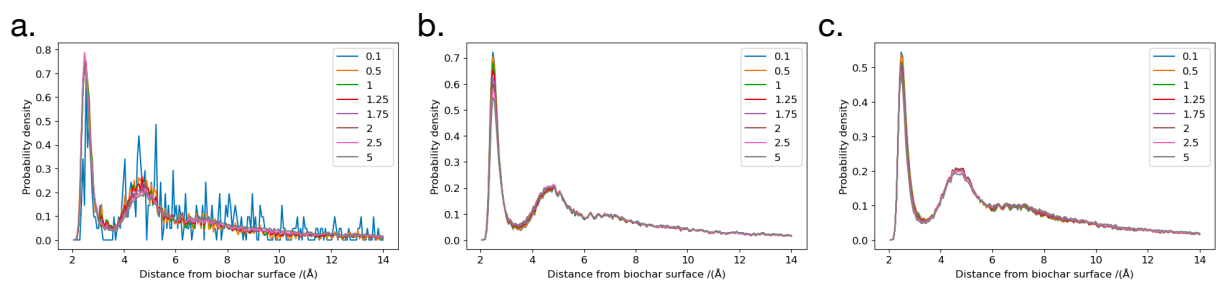

*Figure S3.23*. Distances between Na$^+$ ions and surface-exposed biochar O atoms after equilibration. Each plot shows systems containing a) BC400-surf b) BC600-surf and c) BC800-surf. Starting 2,4-D$^-$ concentration is shown in legend.

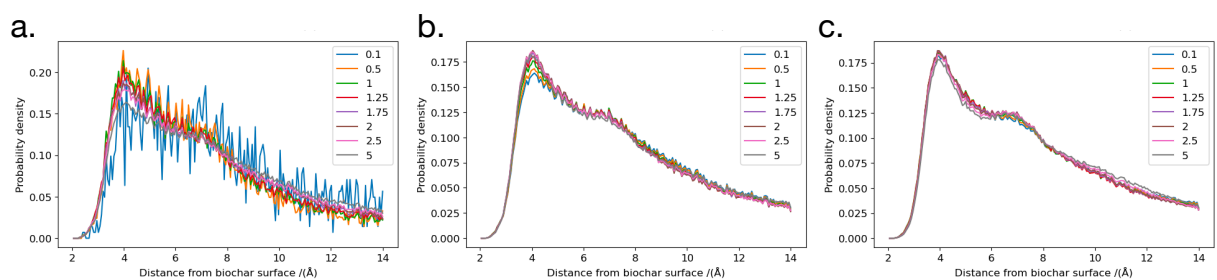

*Figure S3.24*. Distances between Na$^+$ ions and surface-exposed biochar H atoms after equilibration. Each plot shows systems containing a) BC400-surf b) BC600-surf and c) BC800-surf. Starting 2,4-D$^-$ concentration is shown in legend in units of mol L$^{-1}$.

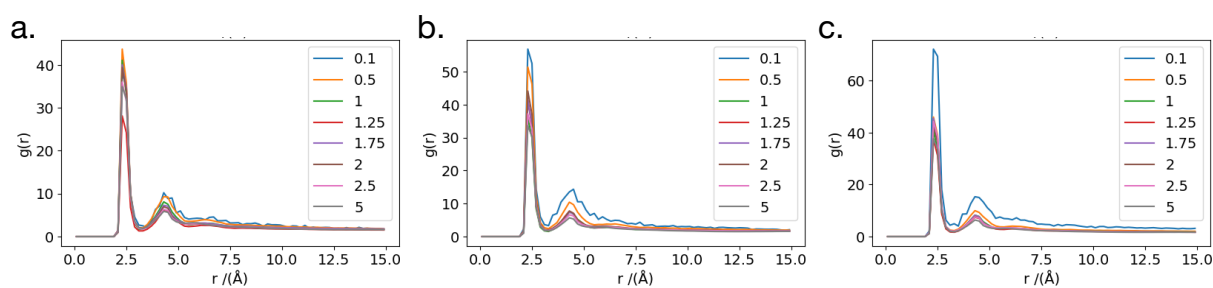

*Figure S3.25*. Radial distribution functions between Na$^+$ ions and 2,4-D$^-$ carboxylate oxygens. Each plot shows systems containing a) BC400-surf, b) BC600-surf and c) BC800-surf. Starting 2,4-D$^-$ concentration is shown in legend in units of mol L$^{-1}$.

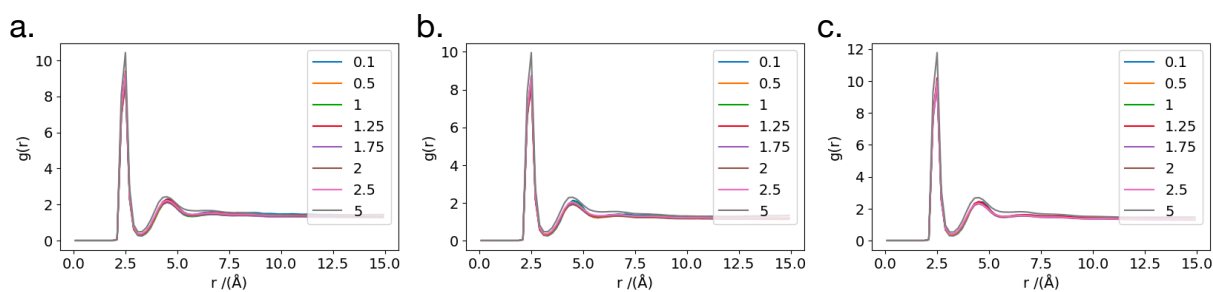

***Figure S3.26***. *Radial distribution functions between $Na^+$ ions and water oxygens. Each plot shows systems containing a) BC400-surf, b) BC600-surf and c) BC800-surf. Starting 2,4-$D^-$ concentration is shown in legend in units of mol $L^{-1}$.*

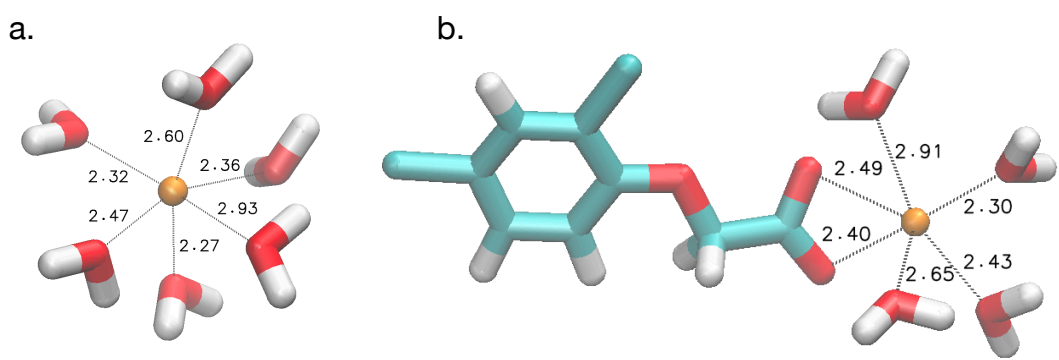

***Figure S3.27***. *Example visualisations of a) an $Na^+$ hydration shell, and b) its partial substitution to form a 2,4-$D^-$ carboxylate oxygens and $Na^+$ complex. Water and 2,4-$D^-$ molecules are coloured in cyan (C and Cl), red (O) and white (H), $Na^+$ ions are coloured in orange. Distances in Å.*

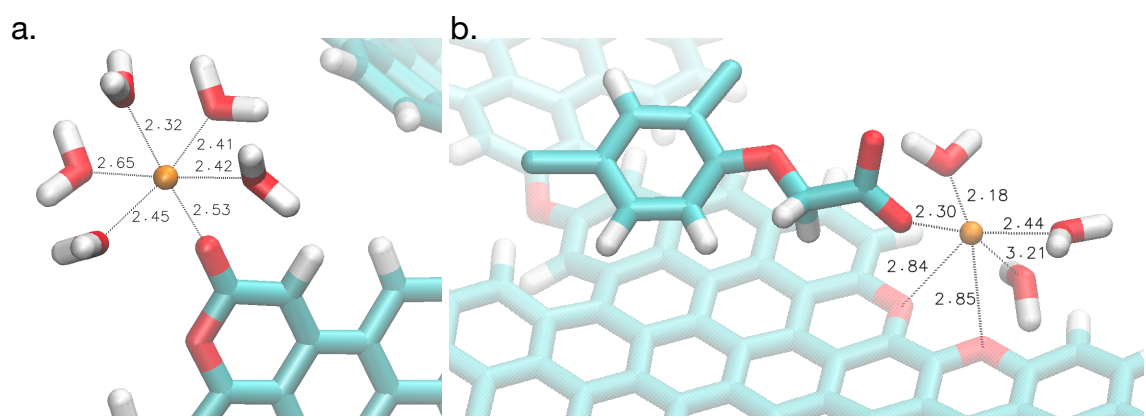

***Figure S3.28***. *Example visualisations of a) the partial substitution of an $Na+$ hydration shell to form a biochar and $Na+$ complex, and b) adsorption of 2,4-$D^-$ via cation bridging. Biochar, water and 2,4-$D^-$ molecules are coloured in cyan (C and Cl), red (O) and white (H), $Na^+$ ions are coloured in orange. Distances are given in Å.*